\documentstyle[12pt,psfig]{article}
\begin{document}

\vskip 1truecm
\rightline{Preprint  PUPT-1649}
\rightline{ e-Print Archive: hep-lat/9610013}
\vspace{0.6in}
\centerline{\Large Curing $O(a)$ Errors in 3-D Lattice 
	                SU(2)$\times$U(1) Higgs Theory}
\vspace{0.6in}
\centerline{\Large Guy D. Moore\footnote{e-mail:
guymoore@puhep1.princeton.edu }
}
\medskip

\centerline{\it Princeton University}
\centerline{\it Joseph Henry Laboratories, PO Box 708}
\centerline{\it Princeton, NJ 08544, USA}

\vspace{0.6in}

\centerline{\bf Abstract}

We show how to make $O(a)$ corrections in the bare parameters of 3-D lattice
SU(2)$\times$U(1) Higgs theory which remove $O(a)$ errors in the match
between the infrared behavior and the infrared behavior of the continuum
theory.  The corrections substantially improve the convergence of lattice 
data to a small $a$ limit.

\smallskip

\begin{verse} 
\end{verse}

\begin{verse}
\end{verse}

\vspace{0.6in}

\section{Introduction}
\label{introduction}

It is believed that the baryon number of the universe may have
originated at the electroweak phase transition.  The details of
the mechanism depend substantially on the details of 
the phase transition itself, so
for several years there has been growing interest in a quantitative 
understanding of the electroweak phase transition.  Ordinary perturbation
theory \cite{Arnold}
has proven useful only when the Higgs mass $m_H \ll m_W$ the
$W$ mass, and outside of this regime the only reliable methods are
nonperturbative, such as lattice Monte-Carlo techniques.

The reason for the failure of perturbation theory was elucidated by the
work of Farakos et. al. \cite{FKRS425,FKRS,KLRS}, who have shown
how, to very good approximation, the thermodynamics of the standard
model near the electroweak phase transition can be described using
a three dimensional path integral for purely bosonic SU(2)$\times$U(1)
Higgs theory.  The theory is super-renormalizable, and hence its
ultraviolet is well behaved, but its infrared is potentially nonperturbative,
leading to the breakdown of perturbation theory as a tool for 
exploring the basically infrared physics of the phase transition.
However, this same feature makes the numerical investigation of the
nonperturbative physics particularly feasible; fine lattices are 
possible because the system is low dimensional, the couplings fall 
quickly with lattice spacing, and there is no need for the (numerically
expensive) inclusion of fermionic degrees of freedom.

Several groups have conducted numerical experiments on the three
dimensional system to study the phase transition 
\cite{several1,several11,several12,several2,several3,several4}.
Each study used the Wilson action and the bare relations 
between lattice couplings and
wave functions and the 3-D continuum parameters.  In each case there
were substantial linear in $a$ (lattice spacing) errors which had to be 
removed by extrapolation over several values of $a$.  The extrapolation
process is numerically expensive because it demands very fine lattices,
and the errors in the final result are dominated by $O(a^2)$ errors
from the coarsest lattice used, and statistical errors in the finest
lattices.  It would be better to have an analytic
understanding of the origin of the linear in $a$ corrections, which
would make it possible to prevent them.  One could then take data
only on lattices of intermediate coarseness, obtaining smaller $O(a^2)$
errors and better statistics with less numerical effort.
This paper presents an analysis of these $O(a)$ errors.

The $O(a)$ errors in measured infrared phenomena arise because the
infrared theory is clothed by interactions with more ultraviolet modes,
but the clothing differs between lattice and continuum theories, both
because the most ultraviolet degrees of freedom are absent on the lattice,
and because the ultraviolet lattice modes have incorrect
dispersion relations and extra (lattice artefact) interactions.  Since
the dispersion relations and couplings of the lattice theory agree with
those of the continuum theory in the infrared, the
difference between clothing in the two theories is all ultraviolet, 
and therefore perturbatively computable.  Also, since the origin of 
the difference is ultraviolet, it should be expressible as an
operator expansion, and at the desired level of accuracy only the 
superrenormalizable terms are needed; these can be compensated for by
a shift in the bare parameters of the theory.  

Dimension two operators, such as the mass squared and the
$\Phi^2$ operator insertion, receive linearly divergent corrections at
one loop; logarithmic corrections, computed in 
\cite{FKRS,Laine}, at two loops; and $O(a)$ corrections
at three loops.  The problem of determining these $O(a)$ corrections is
difficult.  However, if one is only interested in properties of
the phase transition such as jumps in order parameters, the latent
heat, the surface tension, etc. then it is not necessary to know
these to high accuracy.  To compute the jump in $\Phi^2$, for
instance, one only needs to account for the multiplicative correction
to the operator insertion, not the additive correction, which is
common to the two phases.  Similarly, precise knowledge of $m_H^2$ is only
necessary to compute very precisely the equilibrium temperature.
We will not pursue improvement in the determination of these
quantities here, but will concentrate on improving the precision
of the measurement of quantities related to the strength 
of the phase transition, which is more physically interesting.

The outline of the paper is as follows.  In Section \ref{setup} we 
present the problem, establish notation, and give the results.  Readers
uninterested in the details of the calculation can then skip Section 
\ref{details}, which enumerates the required Feynman diagrams, presents
the calculation of a few, and gives results for the others.  The 
conclusion, Section \ref{conclusion}, contains an example of using 
the improvement technique to reinterpret existing data.
The first two appendicies
contain integrals, identities, and Feynman rules needed in Section 
\ref{details}, and the last presents the results for the abelian Higgs
model, and for the standard model but including the adjoint scalar 
($A_0$) fields.

\section{Set-up and Results}
\label{setup}

The dimensional reduction program shows that the infrared thermodynamics of 
the standard model are described up to small error by the 
partition function
\begin{eqnarray}
Z & = & \int {\cal D} \Phi {\cal D} A_i {\cal D} B_i \exp( - \beta 
\int d^3 x H ) \, ,\\
H & = & \frac{1}{4 g^2} \left( F_{ij}^a F_{ij}^a 
+ \frac{1}{\tan^2 \Theta_W} f_{ij} f_{ij} \right) + (D_i \Phi)^{\dagger}
(D_i \Phi) + m_H^2(\mu) \Phi^{\dagger} \Phi 
+ \lambda (\Phi^{\dagger} \Phi)^2 \, .
\end{eqnarray} 
Here $A_i$ and $B_i$ are the gauge fields and 
$F^a_{ij}$ and $f_{ij}$ are the field strengths of the 
SU(2) and U(1) gauge groups respectively, and the couplings and wave
function normalizations are determined with respect to physical measurables
as discussed in \cite{KLRS}.  Only $m_H^2$ depends on the ($\overline{
\rm{MS}}$) renormalization point $\mu$.

The corresponding (Wilson) lattice action, in lattice units, is
\begin{eqnarray}
Z & = & \int {\cal D} U {\cal D} u {\cal D} \Phi
\exp( - \beta_L \sum_x H_L ) \, , \\
H_L & = & Z_A \sum_{i<j} (1 - \frac{1}{2} {\rm Tr} P_{ij}) 
	+ \frac{Z_B}{\tan^2 \Theta_W} \sum_{i<j} ( 1 - {\rm Re} \: p_{ij} )
\nonumber \\ & & 
	+ Z_{\Phi} \left( \sum_i \left( \Phi_a^2(x) 
	- \Phi_a(x) U_{iab}(x) u_{ibc}(x)
	\Phi_c(x+i) \right) + \frac{m_{HL}^2}{2} \Phi_a^2 \right) 
	\nonumber \\ 
	& & + \frac{(\lambda_L - \delta \lambda_L) Z_{\Phi}^2}{4} 
	(\Phi^2_a)^2 \, ,    \label{Wilsonaction}
\end{eqnarray}
where $\Phi$ is treated as four independent real entries $\Phi_a$, 
$U$ and $u$ are the SU(2) and U(1) link variables respectively, and
$P_{ij}$ and $p_{ij}$ are the SU(2) and U(1) $1\times 1$ plaquettes.
The notation for $\Phi_a$ is equivalent to the common notation in which
the Higgs field is written as a magnitude times an $SU(2)$ matrix, and
each sum on $a$ is replaced by $(1/2) {\rm Tr}$.

The (naive) tree relations between the lattice 
parameters and the 3-D continuum ones are
\begin{eqnarray}
\beta_L & \equiv & \frac{4 \beta}{g^2 a}  \, , \label{betal} \\
Z_A & = & 1 \label{ZA} \\
Z_B & = & 1 \label{ZB} \\
Z_{\Phi} & = & 1 \label{Zphi} \\
\lambda_L & \equiv & \frac{4 \lambda}{g^2} \, ,
\quad \delta \lambda_L = 0 \, , \label{lambdal} \\
\Phi_{op}^2 & = & \frac{8 \beta^2
	\Phi^{\dagger} \Phi}{\beta_L^2 g^2} 
	= \frac{ 4 \beta^2 \phi^2_{op}} {g^2 \beta_L^2 } 
	\, , \label{phisquared} \\
m_{HL}^2 & = & \frac{ 16 \beta^2 m_H^2}{g^4 \beta_L^2} \, , 
	\label{mHL}
\end{eqnarray}
where $\Phi^2_{op}$ is the $\Phi^2$ operator insertion.

The super-renormalizability of the theory ensures that, in the limit
$a \rightarrow 0$ (or $\beta_L \rightarrow \infty$), using these
relations--except for Eqs. (\ref{phisquared}) and (\ref{mHL}), 
which must be corrected for one and two loop divergences--will make
the thermodynamics of the lattice system match the thermodynamics
of the continuum theory.  However, the approach to the thermodynamics
of the continuum theory need not be very fast; while we are assured 
that it is power-law, it proves to be $O(\beta_L^{-1})$ (or $O(a)$).
This is because one loop effects introduce $O(\beta_L^{-1})$ corrections
to the righthand sides of Eqs. (\ref{ZA}) through (\ref{mHL}), as 
discussed below.  In the case of Eqs. (\ref{phisquared}) and (\ref{mHL}), 
the $O(\beta_L^{-1})$ correction swamps the $O(\beta_L^{-2})$ tree
value; there is a linear divergence, which should be tamed in order to
control the lattice theory.  This correction, and 2 loop $O(\beta_L^{-2} 
\ln \beta_L)$ corrections, are computed in \cite{FKRS,Laine}.  The 
corrections to the other equations have not previously been computed.

To see how $O(\beta_L^{-1})$ corrections arise in the infrared behavior,
consider some Feynman diagram needed in the evaluation of some 
infrared quantity of interest, such as the effective potential
or the domain wall surface tension.  The dominant contribution arises
when all loop momenta are on order the nonperturbative scale $p \sim g^2 T$,
or in lattice units $p \sim \beta_L^{-1}$.  At these momenta the theories
match except for $O(\beta_L^{-2})$ nonrenormalizable operators, which
are not important at the desired level of accuracy.  There
are also contributions where one or more loop momenta are ``hard,''
$p \sim 1$ in lattice units.  At such momenta the lattice and continuum
theories differ significantly.  However, at these momenta the
theory is perturbative, by the same power counting argument which 
establishes its super-renormalizability.  For the wave function
normalizations of Eq. (\ref{Wilsonaction}), each vertex contributes a
power of $\beta_L$ and each propagator contributes a power of 
$\beta_L^{-1}$, so the result is suppressed parametrically by one power
of $\beta_L^{-1}$ for each hard loop.  The value of a hard loop with all
soft incoming momenta can be computed once and for all in an expansion in
the momenta of the incoming soft lines, so the difference between the
lattice and continuum hard loops can be written as an operator product
expansion.  Only dimension 2 and 4 (super-renormalizable) operators
will be important at $O(\beta_L^{-1})$ for the evaluation of the remaining
soft diagram, as higher dimension operators just contribute small 
corrections to the nonrenomalizable operators which could already be
neglected.  And the super-renormalizable operators can be absorbed
by $O(\beta_L^{-1})$ counterterms in the tree level Lagrangian,
making the remaining soft integrations match between the lattice and
continuum theories, up to $O(\beta_L^{-2})$ corrections.  In other
words, we resum the class of all diagrams with non-overlapping hard loops
by collapsing the hard loops into operator insertions and applying 
counterterms in the Lagrangian to absorb the difference between the
lattice and continuum values.  The lattice and continuum perturbation
series then match up to $O(\beta_L^{-2})$ corrections when we sum over
the subset of diagrams with isolated hard loops, rather than just
those diagrams without hard loops, which is the subset for which 
tree relations between lattice and continuum parameters lead to 
matching contributions.  Since overlapping hard loops
are suppressed by two powers of $\beta_L^{-1}$, by the same power counting
used earlier, the infrared theories will now match at $O(\beta_L^{-1})$.

An alternate way of phrasing the same argument is to consider how the
lattice theory can arise from integrating out degrees of freedom from the 
continuum theory.  We want the infrared behavior to match when we 
integrate out these degrees of freedom.  At tree level one just relates
the couplings and wave functions by naive dimensional analysis.  But to
achieve higher accuracy one must take into account the loop effects
which were provided by the degrees of freedom no longer present, and the
change in loop effects due to the inevitable changes in the theory at
the lattice scale such as the new lattice dispersion relations and 
added (lattice artefact) couplings.  The influence of these loop effects
on the infrared physics should be computed; as discussed above it can be
expressed at the desired level of accuracy as perturbatively small 
operator insertions, of which only the super-renormalizable ones are
relevant.  One should then shift the values of the lattice theory
Lagrangian parameters to absorb the difference between the lattice and
continuum contributions, ie one replaces loop effects which would appear
in the continuum theory but which are absent in the lattice theory
with shifts in the lattice parameters which reproduce these effects.
The corrections are perturbative with perturbation expansion parameter
$\beta_L^{-1}$, because they all arise from momentum scales on order
or greater than the reciprocal lattice spacing, and the power counting
discussed above applies.  (The only exception would be if there were
infrared divergences, but since the infrared lattice and continuum theories
match, the difference in loop contributions between the two theories
will be free of infrared divergences.)  At the desired level of 
accuracy we need to perform a one loop match.

A complication is dimension 2 operators.  The most general renormalization
of these operators is
\begin{eqnarray}
m^2_{HL,{\rm renormalized}} & = & Z_m \frac{16 \beta^2 m^2_H}{g^4 \beta_L^2}
	+ \delta m^2_{HL} \, , \label{mHrenorm} \\
\Phi_{op,{\rm renormalized}}^2 & = & Z_{op} \frac{ 4 \beta^2 
	\phi^2_{op}} {g^2 \beta_L^2 } + \delta \Phi_{op}^2 
	\, .\label{phirenorm}
\end{eqnarray}
In addition to a multiplicative renormalization, $Z_m$ and $Z_{op}$, 
there is an additive renormalization, $\delta m_{HL}^2$ and
$\delta \Phi^2_{op}$.  The problem is that in lattice units the tree
level $m_{HL}^2$, and the expected value of $\Phi_{op}^2$ due to the
interesting infrared physics, is $O(\beta_L^{-2})$.  The $O(\beta_L^{-1})$
corrections $\delta m_{HL}^2$ and $\delta \Phi_{op}^2$ will swamp
these.  That is why it was considered necessary to carry the calculation
of these counterterms to two loops.  To determine the counterterms to
$O(\beta_L^{-1})$ accuracy relative to the tree value would require a
prohibitively hard three loop calculation.  However, in practice we only
need to know $\delta m_{HL}^2$ with precision if we want to determine the 
phase transition temperature with high accuracy, since the relation
between the 3 dimensional continuum value of $m_H^2$ and the physical, 
4 dimensional Higgs mass squared is strongly temperature dependent.  In
practice one tunes $m_{HL,ren}^2 - \delta m_{HL}^2$ to find the phase
transition temperature; to find the absolute temperature this represents,
one must know $\delta m_{HL}$ with precision, but to determine 
the splitting between the equilibrium temperature and the spinodal 
temperatures, for instance, one only needs to know $Z_m$; only this 
quantity is relevant for details of the strength of the phase transition.
Similarly, we need to know $\delta \Phi^2_{op}$ to find the absolute
value of $\phi^2$ in one phase; but the question which is more interesting
physically is the difference in $\phi^2$ between the two electroweak
phases, and $\delta \Phi^2_{op}$ cancels out in such differences, and only
$Z_{op}$ is relevant.  Furthermore, because the mass squared 
and the $\Phi^2$ operator insertion couple to the same operator, $Z_m
= Z_{op}$; so henceforth we will only consider $Z_{op}$ (though we will
also reproduce the (known) one loop value for $\delta m_{HL}^2$ as a
natural biproduct of our renormalization calculation).

\begin{figure}
\centerline{\psfig{file=fig1.epsi,width=\hsize}}
\caption{\label{fig1}  Relation between bare and clothed propagator at
one loop.}
\end{figure}

Now we proceed to set up the computation of the renormalization constants
needed in the match between continuum and lattice theories.  We begin
with the scalar propagator. 
In the continuum theory the value of the scalar propagator is 
$G(p^2) = 1/( p^2 + m_H^2 - \pi_{\Phi,C}(p^2))$, 
with $\pi_{\Phi,C}$ the continuum theory self-energy
computed using the clothed vertices and propagators, expressed
in lattice units.  In the lattice
theory the bare propagator is modified by $Z^{-1}_{\Phi}$, so 
the clothed propagator is (see Figure \ref{fig1})
\begin{equation}
G = \frac{Z^{-1}_{\Phi}}{p^2+ m_{HL}^2 - \delta m_{HL}^2} 
	\left( 1 + \pi_{\Phi,L} G \right) \, ,
\end{equation}
and hence
\begin{equation}
(Z_{\Phi} - 1) p^2 + (Z_{\phi}-1) m_{HL}^2- Z_{\Phi} \delta m_{HL}^2 
= \pi_{\Phi,L} - \pi_{\Phi,C} \, .
\end{equation}
Hence one can determine $\delta m_{HL}^2$ and $Z_{\Phi}$ from
the difference between the lattice and continuum self-energies at
$O(p^0)$ and $O(p^2)$.  The self-energies
should be computed using clothed vertices and wave functions, which 
coincide (by hypothesis) in the infrared; so the difference will be 
ultraviolet dominated, and here the clothed vertices and wave functions
can be replaced by bare ones (without $Z$ and $\delta \lambda$ corrections)
with an $O(\beta_L^{-1})$ error, which
would be accounted for in a full two loop calculation and will only lead
to an $O(\beta_L^{-2})$ error in $Z_{\Phi}$, and can therefore be
neglected.  Similarly, the $(Z_{\Phi}-1)m_{HL}^2$
can be dropped ($m^2_{HL}$ being $O(\beta_L^{-2})$), 
and the $Z_{\Phi}$ in front of $\delta m_{HL}^2$ can
be set to 1.

Next, denoting the loop contribution to the amputated 1PI four
point function at small (zero) external momentum
as $-V_{\lambda}$ (the minus sign because the contribution
from the scalar self-coupling is $-\lambda$), and again denoting
the lattice and continuum values with $L$ and $C$ subscripts
respectively, one demands that the actual strength of the scalar
self coupling coincide between lattice and continuum theories, 
\begin{equation}
(\lambda_L - \delta \lambda_L) Z_{\Phi}^2 + V_{\lambda,L} = \lambda_L
+ V_{\lambda,C} \, ,
\end{equation}
and hence, at leading order in $\beta_L^{-1}$,
\begin{equation}
\delta \lambda_L = V_{\lambda,L} - V_{\lambda,C} 
	+ 2 \lambda ( Z_{\Phi} - 1 ) \, .
\end{equation}
Again, the $V_{\lambda}$ are computed with the clothed vertices and
propagators, but can be computed using the naive bare propagators
and vertices because the difference will be ultraviolet dominated.  

\begin{figure}
\centerline{\psfig{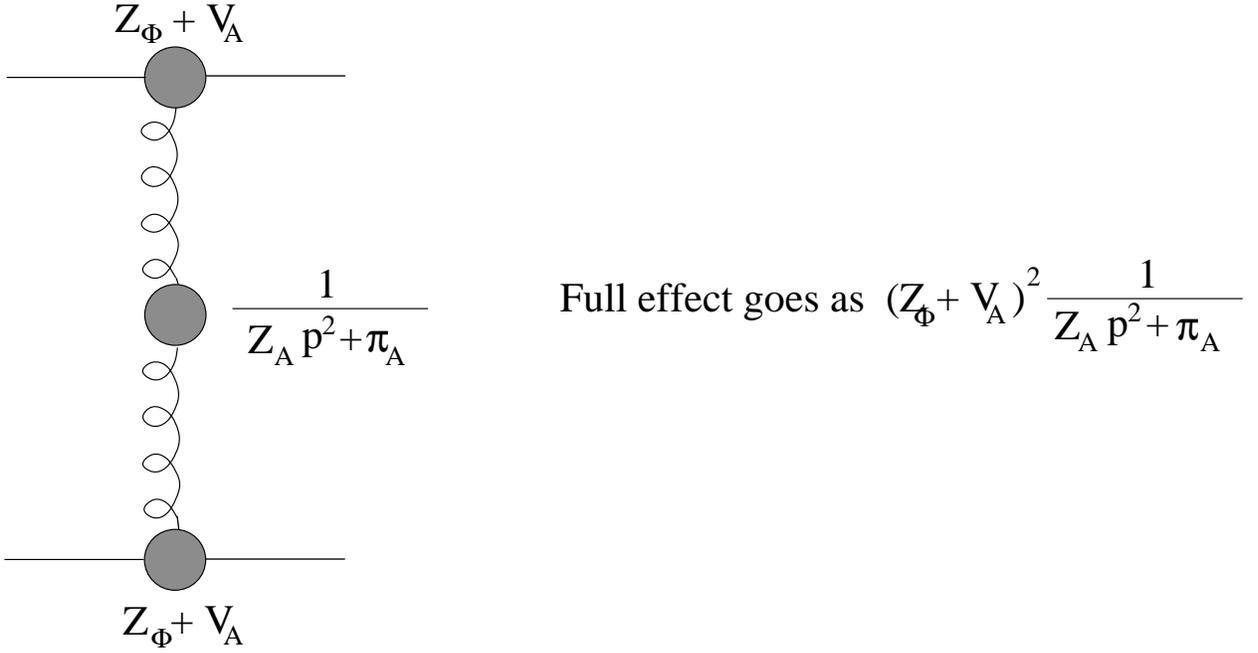}}
\caption{\label{gaugefix} The full influence of a gauge particle
propagating between two lines depends on the vertex and propagator
renormalizations.  Here and throughout SU(2) lines will be curly and
U(1) lines will be wiggly.}
\end{figure}

To find $Z_A$ and $Z_B$ one must compute not only the gauge field
self-energy corrections $\pi_A$ and $\pi_B$, but also the corrections
to the gauge-scalar vertex, which we will denote as $V_A$, $V_B$.  (It does
not matter which vertex is considered, by gauge invariance; we
choose the scalar vertex because it is the easiest to compute.)  $Z_A$ should
be chosen so that the complete effect of a gauge particle propagating
between two lines, say two scalar lines, is the same between the lattice
and continuum theories, see Figure \ref{gaugefix}.  
The strength of the vertex at each end
of the propagator is multiplied by
$Z_{\Phi} + V_{A,L}$  in the lattice theory and $1 + V_{A,C} $ in the
continuum theory; the $Z_{\Phi}$ factor arises because its appearance in
the action
modifies both the wave function and the coupling to the gauge field.  The
modification of the propagator is similar to that for the scalar,
and we find that to match the full modification (a vertex at each end
and a propagator in the middle) we must choose $Z_A$ such that
\begin{equation}
\frac{ ( Z_{\Phi} + V_{A,L} )^2}{Z_A p^2 - \pi_{A,L}} = 
\frac{1+ V_{A,C}}{p^2 - \pi_{A,C}} \, ,
\end{equation}
or, at leading order in $\beta_L^{-1}$, 
\begin{equation}
Z_A - 1 = 2 (Z_{\Phi} - 1) + 2 ( V_{A,L} - V_{A,C} )
	+ \frac{\pi_{A,L} - \pi_{A,C}}{p^2} \, .
\end{equation}
The equation for $Z_B$ is analogous, and in this case the vertex and
Higgs wave function contributions will turn out to cancel.  

The $\Phi^2$ operator multiplicatively renormalizes due to the diagrams
shown in Figure \ref{Op_renorm}, because in every diagram where the operator 
appears on an infrared Higgs field line, there is a corresponding 
diagram where the $\Phi^2$ insertion on that line is replaced with
the one loop insertion shown in that figure.
If the sum of the contributions of the diagrams there,
at small (zero) external momentum, are
denoted by $L_{\Phi,L}$, and the continuum contribution is $L_{\Phi,C}$,
then the renormalization of the $\Phi^2$ operator is
\begin{equation}
Z_{OP} = 1 - ( L_{\Phi,L} - L_{\Phi,C} ) \, ,
\end{equation}
where each $L$ should be computed with clothed propagators,
vertices, and operator insertions.  Again, at the correct value of 
$Z_{OP}$ the infrared definition of the operator insertions will coincide,
so there is no infrared divergent contribution to the difference;
the difference is UV dominated and can be computed, at 
the desired level of accuracy, with the bare vertices and propagators.

It is possible to scale $\Phi$ by a constant $\kappa$ without changing 
the physics, provided one multiplies $Z_{\Phi}$ by $\kappa^{-2}$ and
$Z_{OP}$ by $\kappa^{-2}$.  Only the combination $Z_{\Phi} Z_{OP}^{-1}$
must be meaningful and gauge invariant.  It is convenient to perform
such a scaling of $\Phi$ to set $Z_{OP}$ to 1, in which case
one should add $(L_{\Phi,L} - L_{\Phi,C})$ to $Z_{\Phi}$ at the 
very end of the calculation.

The diagrams needed for the renormalization are presented in the next
section, and their values are tabulated there.  All but two are performed in
lattice Lorentz gauge with a general gauge parameter, to check gauge
parameter independence; the exceptions are the gauge field self-energy
diagrams with all gauge vertices, which are only performed in Feynman
gauge; so we have not yet checked the gauge invariance of $Z_A$.
The results of the calculation are
\begin{eqnarray}
Z_A - 1 & = & \beta_L^{-1} \left( \frac{73}{3} \frac{\xi}{4 \pi} 
	+ \frac{1}{12}
	\frac{\Sigma}{4 \pi} + \frac{1}{3} \right) \, , \label{Zaeq} \\
Z_B - 1 & = & \beta_L^{-1} \tan^{2} \Theta_W 
	\left( - \frac{1}{3} \frac{\xi}{4 \pi} + \frac{1}{12}
	\frac{\Sigma}{4 \pi} + \frac{1}{3} \right) \, , \\
Z_{\Phi} Z_{OP}^{-1} - 1 & = & \beta_L^{-1} \left( 
	( 9 + 3 \tan^2 \Theta_W - 6 \lambda ) \frac{\xi}{4\pi}
	+ \frac{3 + \tan^2 \Theta_W}{6} \frac{\Sigma}{4\pi} \right) \, , \\ 
\delta m_{HL}^2 & = & \beta_L^{-1} (6 + 2 \tan^2 \Theta_W + 6 \lambda_L)
	\frac{\Sigma}{4 \pi} \, , \\
\delta \lambda_L & = & \beta_L^{-1} \left( \left( -12\lambda_L^2 - 4
	- 2 (1 + \tan^2 \Theta_W )^2 + 6 \lambda_L (3 + 
	\tan^2 \Theta_W ) \right) \frac{\xi}{4 \pi}  
	+ \right. \label{lambdaeq}\\
	& & \left. \frac{ 3 + \tan^2 \Theta_W}{3} \lambda_L  
	\frac{\Sigma}{4 \pi} \right) \, . \nonumber
\end{eqnarray}
The constants $\Sigma = 3.175911536$ and $\xi = 0.152859325$ have the
same meaning as in \cite{FKRS}, and arise from integrals presented
in Appendix A.  If one is interested in the theory without the U(1)
gauge group, set $\tan^2 \Theta_W = 0$ and disregard $Z_B$.
Equations (\ref{Zaeq})--(\ref{lambdaeq}) are the main result 
of this paper.

By applying these corrections to the Hamiltonian of a lattice simulation
one should be able to remove all $O(a)$ errors (except in the absolute
determination of the order parameter and the
phase transition temperature, as discussed above).  
If one already has numerical data, then the $O(a)$ errors can be removed
by re-interpreting the meaning of the answers.  In the case of SU(2)
Higgs theory alone, equating
\begin{eqnarray}
\beta_{naive} \bigg( \sum 1 - \frac{1}{2} {\rm Tr} P_{ij} + \frac{1}{2}
	\Phi_{a,naive} D^2_{Latt,ab} \Phi_{b,naive} +
\nonumber \\ 
	\frac{m_{HL}^2 - \delta m_{HL}^2}{2} \Phi^2_{naive}
	+ \frac{\lambda_{L,naive}}{4} (\Phi^2_{naive})^2 \bigg)
\end{eqnarray}
with
\begin{eqnarray}
\beta_{imp} \bigg( Z_A \sum 1 - \frac{1}{2} {\rm Tr} P_{ij} + \frac{1}{2}
	Z_{\Phi} \Phi_{a,imp} D^2_{Latt,ab} \Phi_{b,imp} +
\nonumber \\
	 Z_{\Phi} \frac{m_{HL}^2 - \delta m_{HL}^2}{2} \Phi^2_{imp}
	+ Z_{\Phi}^2 \frac{\lambda_{L,imp} - \delta \lambda_L}{4} 
	(\Phi^2_{imp})^2 \bigg) \, ,
\end{eqnarray}
one finds as matching conditions 
\begin{eqnarray}
\beta_{imp} & = & Z_A^{-1} \beta_{naive} \, , \label{impbeta} \\
\lambda_{L,imp} & = & Z_A^{-1} \lambda_{L,naive} 
	+ \delta \lambda_L \, , \label{implam} \\
\beta^2_{imp} Z_{OP} \Phi^2_{imp} & = & Z_{\Phi}^{-1} Z_{OP} 
	Z_A^{-1} \beta_{naive}^2
	\Phi^2_{naive} \, . \label{impphi}
\end{eqnarray}
One should re-interpret at what temperature the simulation was conducted,
what input value of $\lambda_L$ was used, and what the result for 
$\beta_L^2 \Phi^2$ was\footnote{If we include the U(1) subgroup, this
procedure does not work quite as well; the value of $\tan^2 \Theta_W$ will
differ before and after improvement, which might not be desirable.}.  
Note both that the value of $\lambda_{L,imp}$ is
smaller than the value of $\lambda_{L,naive}$ and that the rescaling
of $\beta_L^2 \Phi^2$ reduces its value.  Both lead to an overestimate
(before correction) of the strength of the phase transition, the first
because the phase transition is stronger at smaller $\lambda$ and
the second because the real jump in the order parameter is smaller 
than that using naive conversions.  The error in $\lambda$ is the most
problematic, because at small values of $\lambda$ 
the strength of the phase transition
is strongly $\lambda$ dependent (one loop perturbation theory, which
applies here, says the jump in $\Phi^2$ should go as $\lambda^{-2}$),
and at larger values of $\lambda$ the properties of the phase
transition may change qualitatively as one changes $\lambda$.  Note also
that $Z_A - 1 \simeq 0.65 \beta_L^{-1}$ is quite large, and so are 
corrections where $Z_A$ appears.

One can get a rough understanding of the magnitudes of $Z_A$, $Z_B$, 
$Z_{\Phi}$, and $\delta m_{HL}^2$ by considering the tadpole correction
scheme of Lepage and Mackenzie \cite{Lepage}.  Based on a mean field 
theory argument, they propose that the dominant loop contributions at
every order can be absorbed by making an ``educated guess'' for the
constants $Z_A$, etc, as follows: compute the mean value of the
trace of an elementary plaquette, and call it $U$ (or $u$ for the U(1)
plaquette).  Then guess that every term in the action which contains
a link should be multiplied by $U^{-1/4}$ (with the value of $U$ found
after the correction has been applied, so $U$ must be found
self-consistently).  Hence, $Z_A$ should 
roughly equal $U^{-1}$, $Z_{\Phi}$ should roughly equal $U^{-1/4}u^{-1/4}$,
$Z_B$ should roughly equal $u^{-1}$, and $\delta m_{HL}^2$ should
be about $6(U^{-1/4}u^{-1/4}-1)$ (because the hopping term in
the Higgs wave function gets corrected, but not the local term).
It is quite easy to compute at one loop that $U = 1 - \beta_L^{-1}$
and $u = 1 - \tan^2 \Theta_W \beta_L^{-1}/3$.  
(This is just the approximation that
the average energy in the plaquette term in the Hamiltonian obeys 
equipartition.)  Hence, one guesses that, at lowest order in $\beta_L$,
\begin{eqnarray}
\beta_L ( Z_A - 1 ) & \simeq & 1 \, , \\
\beta_L ( Z_B - 1 ) & \simeq & \frac{\tan^2 \theta_W}{3} \, , \\
\beta_L ( Z_{\Phi}Z_{OP}^{-1} - 1 ) & \simeq & 
	\frac{3 + \tan^2 \Theta_W}{12} \, , \\
\beta_L \delta m^2_{HL} & \simeq & \frac{ 3 + \tan^{2} \Theta_W}{2} \, .
\end{eqnarray}
In fact, at $\lambda_L = 0$, the numerical values of the corrections are
\begin{eqnarray}
\beta_L ( Z_A - 1 ) & = & 0.65 \, , \\
\beta_L ( Z_B - 1 ) & = & 0.350 \tan^2 \Theta_W \, , \\
\beta_L ( Z_{\Phi}Z_{OP}^{-1} - 1 ) & = & 
	0.0786 (3 + \tan^2 \Theta_W) \, , \\
\beta_L \delta m^2_{HL} & = & 0.505 ( 3 + \tan^2 \Theta_W ) \, ,
\end{eqnarray}
which are all close.  The tadpole argument is least accurate for the
SU(2) wave function correction $Z_A - 1$.

The tadpole improvement value for the correction
to the scalar self-coupling is that it should only change due to the 
wave function correction.  This misses the corrections at even powers
of $\lambda_L$ (though these are all suppressed by the rather small number 
$\xi/(4\pi)$).  These corrections are important when $\lambda_L$
is fairly small, because results are very $\lambda_L$ sensitive in 
this regime.  This illustrates a limitation of the tadpole improvement
scheme in a theory with non-gauge couplings.

\section{Details of the calculation}
\label{details}

Here we will present the calculations of the vertices and self-energies
needed in the last section in more detail.  To illustrate what is involved
in the calculation
we will present the full details of the contribution of loops involving
gauge particles to the scalar self-energy and of scalar loops to
the gauge field self-energy.  For all other
diagrams we will simply present results. 

\begin{figure}
\centerline{\psfig{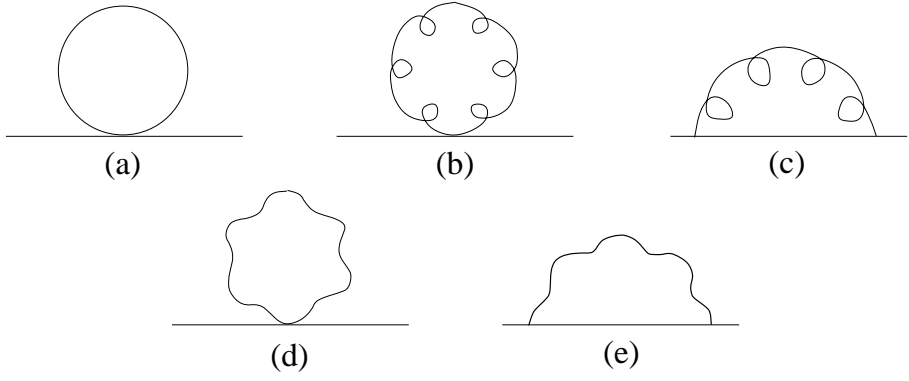}}
\caption{\label{Higgs_self} Diagrams contributing to the Higgs boson
self-energy.  The curly line refers to the SU(2) gauge particle,
the wiggly line to the U(1) gauge particle, and the solid line to
the scalar.}
\end{figure}

The diagrams contributing to
the Higgs self-energy are presented in Figure \ref{Higgs_self}. Diagram
$(a)$ is computed already in \cite{FKRS}.  It is straightforward and leads
to an $O(p^0)$ contribution to the difference in self-energies of
\begin{equation}
\pi_{\Phi,L} - \pi_{\Phi,C} = \beta_L^{-1} \left( -6 \lambda_L 
	\frac{\Sigma}{4 \pi} \right) \qquad {\rm from} \; (a) \, .
\end{equation}

Diagram $(b)$ vanishes in the $\overline{\rm MS}$ regulated continuum
theory, so its contribution to the difference of self-energies in
Landau gauge is
\begin{equation}
- \frac{3}{\beta_L} \int_{\left[-\pi,\pi\right]^3} 
	\frac{d^3 k}{(2 \pi)^3} \frac{\delta_{ij} 
	\left(\delta_{ij} - \frac{\tilde{k}_i \tilde{k}_j}{\tilde{k}^2} 
	\right)}{\tilde{k}^2} \cos p_i \, ,
\end{equation}
where here and throughout $p$ is the momentum of the line, 
$\tilde{k}_i = 2 \sin(k_i/2)$, and $\tilde{k}^2 = \sum_{i=1,2.3} 
\tilde{k}_i^2$.  The factor of $3$ is a group factor.

To make expressions more concise, the factor $d^3 k/(2\pi)^3$ will
henceforth be implied in every integral, continuum or lattice.

At small $p$, which is the interesting regime, 
$\cos(p_i) \simeq 1 - p_i^2/2$, and the contribution from diagram $(b)$
becomes
\begin{equation}
\frac{-3}{\beta_L} \int_{\left[-\pi,\pi\right]^3}  \left(
	\frac{2}{\tilde{k}^2}
	- \sum_i \frac{p_i^2}{2} \frac{ \tilde{k}^2 - \tilde{k}_i^2}
	{(\tilde{k}^2)^2} \right) \, .
\end{equation}
The integral of $\tilde{k}_i^2/(\tilde{k}^2)^2$ is just $1/3$ the integral
of $1/\tilde{k}^2$ by cubic invariance, so the contribution of the
diagram is 
\begin{equation}
\pi_{\Phi,L} - \pi_{\Phi,C} = \beta_L^{-1} ( -6 + p^2 ) \frac{\Sigma}
	{4 \pi} \qquad {\rm from} \; (b) \, .
\end{equation} 
The contribution of $(d)$ differs only in counting factors; the $3$
is replaced with $\tan^{2} \Theta_W$, which arises from the U(1) 
propagator.

In Landau gauge, diagram $(c)$ on the lattice contributes
\begin{equation}
\pi_{\Phi,L} \; {\rm from} \; (c) \; = 
	\frac{3}{\beta_L} \int_{\left[-\pi,\pi\right]^3} \frac{ 
	(\widetilde{2p-k})_i (\widetilde{2p-k})_j
	\left( \delta_{ij} - \frac{ \tilde{k}_i \tilde{k}_j}{\tilde{k}^2}
	\right)}{\tilde{k}^2 (\widetilde{p-k})^2} \, .
\end{equation}
The continuum contribution is
\begin{equation}
\pi_{\Phi,C} \; {\rm from} \; (c) \; = \frac{3}{\beta_L} \int_{\Re^3}
	\frac{(2p-k)_i (2p-k)_j
	\left( \delta_{ij} - \frac{k_i k_j}{k^2} \right)}
	{k^2 (p-k)^2} \, .
\end{equation}
Each numerator vanishes at $p=0$, so to extract the $O(p^2)$ behavior we
need only expand the numerator to $O(p^2)$ and use the denominator
at $p=0$.  (Each contribution is then separately infrared divergent, but
the infrared divergences match and do not matter to the difference.)
Since
\begin{equation}
(\widetilde{2p-k})_i = 2 \sin(p_i-k_i/2) = -\tilde{k}_i \cos(p_i) + 
\widetilde{2p}_i \cos(k_i/2) \, ,
\end{equation}
and $\tilde{k}_i$ annihilates against the gauge propagator (only because
we are in Landau gauge), the contributions simplify at small $p$ to
\begin{eqnarray}
\pi_{\Phi,L} - \pi_{\Phi,C} \; {\rm from} \; (c) \; & = & \frac{3}{\beta_L}
	 \sum_{ij} 4 p_i p_j \left( \int_{\left[-\pi,\pi\right]^3} 
	\frac{ \cos(k_i/2) \cos(k_j/2)
	(\delta_{ij} - \tilde{k}_i \tilde{k}_j/\tilde{k}^2)}{(\tilde{k}^2)^2}
	- \right. \nonumber \\ & & \left. \int_{\Re^3} 
	\frac{\delta_{ij} - k_i k_j/k^2}{k^4}
	\right) \, .
\end{eqnarray}
Each integral vanishes when $i \neq j$.  When $i=j$,
one may use $\cos^2(k_i/2) = 1 - \tilde{k}_i^2/4$ and cubic invariance
to reorganize the contribution as
\begin{equation}
\pi_{\Phi,L} - \pi_{\Phi,C} \; {\rm from} \; (c) = 
	\frac{3 p^2}{\beta_L} \left[ \frac{8}{3} 
	\left( \int_{\left[-\pi,\pi\right]^3} 
	\frac{1}{(\tilde{k}^2)^2} - \int_{\Re^3} \frac{1}{k^4} \right)
	- \frac{1}{3} \int_{\left[-\pi,\pi\right]^3} \frac{1}{\tilde{k}^2}
	+ \int_{\left[-\pi,\pi\right]^3} \frac{\tilde{k}_1^4}{(\tilde{k}^2)^3} 
	\right] \, .
\end{equation}

The first two integral expressions here are the constants $\xi/4\pi$ and
$\Sigma/4\pi$.  The last is related to these constants by an identity,
Eq. (\ref{4by6}).  The result is 
\begin{equation}
\pi_{\Phi,L} - \pi_{\Phi,C} \; {\rm from} \; (c) \; = \frac{3 p^2}{\beta_L}
	\left( 3 \frac{\xi}{4 \pi} - \frac{1}{6} \frac{\Sigma}{4\pi}
	\right)
\end{equation}
in Landau gauge.  To get the contribution from diagram $(e)$, replace
the group factor, $3$, with $\tan^2 \Theta_W$.

In this gauge, the numerical value of the contribution to the
wave function renormalization from diagram
$(c)$ is $-0.017/\beta_L$, while the contribution from the ``tadpole''
diagram $(b)$ is $+0.253/\beta_L$.  The tadpole contribution is much
bigger, even though diagram $(b)$ vanishes in the continuum theory and 
would not naively be expected to contribute to the wave function 
renormalization at all.  The contribution of this diagram almost
exactly equals the expectation of the ``tadpole improvement'' 
argument.

For a general gauge parameter, where the gauge propagator becomes
\begin{equation}
\frac{ \delta_{ij} + (\alpha - 1) \frac{\tilde{k}_i \tilde{k}_j}
	{\tilde{k}^2}}{\tilde{k}^2} \, ,
\end{equation}
the contribution from $(b)$ changes by $\alpha (\Sigma/4\pi) (-3 + p^2/2)$,
and the contribution from $(c)$ changes by 
$-\alpha (\Sigma/4\pi) (-3 + p^2/2) - 3 \alpha p^2 \xi/4\pi$.  The
mass squared correction is not gauge parameter dependent, but the
wave function correction is.  This will be matched by a similar
gauge parameter dependence in $Z_{OP}$, so the combination 
$Z_{\Phi} Z_{OP}^{-1}$ is gauge invariant.

\begin{figure}
\centerline{\psfig{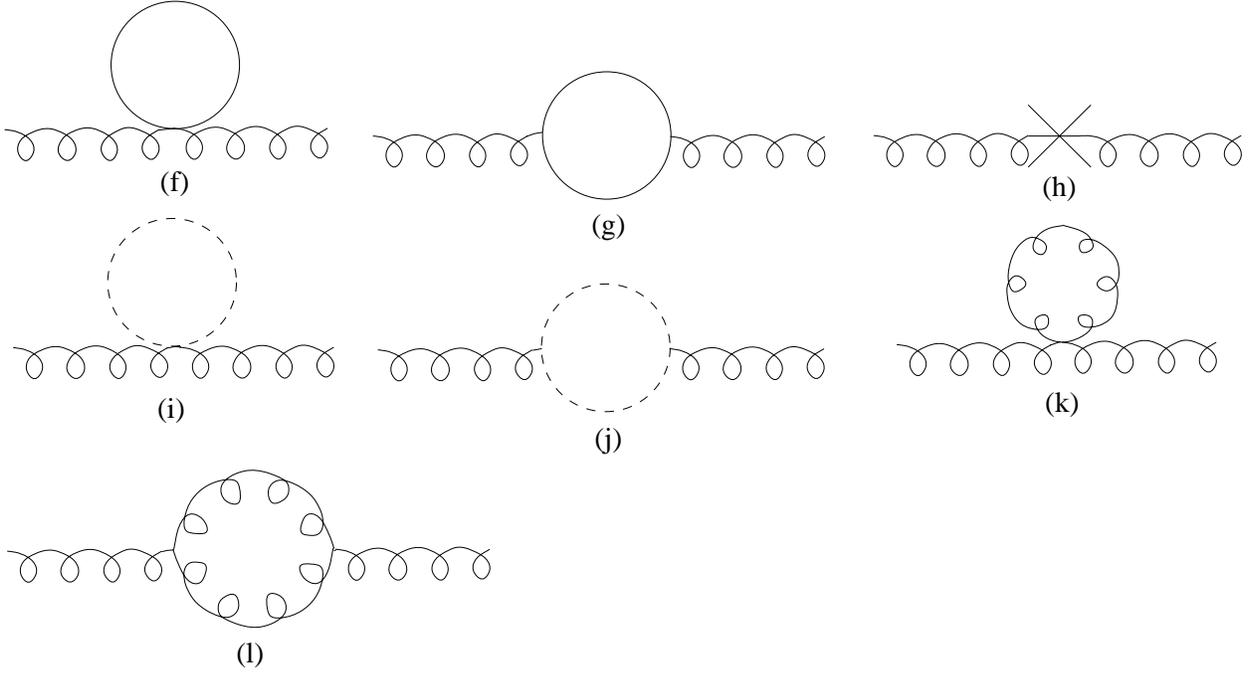}}
\caption{\label{SU2_self} Contributions to the SU(2) gauge particle
self-energy $\pi_A$.  The curly line is the gauge particle, the
solid line is the scalar, and the dotted line is the ghost.}
\end{figure}

New complications arise in computing the gauge particle self-energy.
The value of the self-energy must be transverse and rotationally
invariant, and it must approach zero at $p \rightarrow 0$;
but individual diagrams need not separately satisfy these requirements.
It is a nontrivial check on the calculation if they do, and on the lattice
that check will involve the use of identities, which are intimately
related to gauge invariance.  To illustrate some of these issues we will
present the calculation of the self-energy corrections from scalar loops.
Two diagrams contribute, diagrams $(f)$ and $(g)$ in Figure \ref{SU2_self}.
The contribution of $(f)$, which vanishes in the $\overline{\rm MS}$
renormalized continuum theory, is
\begin{equation}
\pi_{A,L} \; {\rm from} \; (f) = -\frac{2}{\beta_L} \delta_{AB} \delta_{ij} 
	\int_{\left[-\pi,\pi\right]^3} \frac{\cos k_i}{\tilde{k}^2} \, ,
\label{firstp0}
\end{equation}
where the 2 arises from the trace over the fundamental representation
(in the standard continuum notation it would be 1/2, but here the 
coupling involves the Pauli matrix $\tau$ rather than $\tau/2$), 
and the $\delta_{AB}$ is in
group space.  Using $\cos k_i = 1 - \tilde{k}_i^2/2$ and cubic invariance,
we perform the integrals and get
\begin{equation}
\pi_{A,L} \; {\rm from} \; (f) = -\frac{2}{\beta_L} \delta_{AB} \delta_{ij} 
	\left( \frac{\Sigma}{4\pi} - \frac{1}{6} \right) \, .
\end{equation}
This does not vanish as $p \rightarrow 0$, and in fact does not depend
on $p$ at all.  It is also independent of the gauge parameter $\alpha$,
because no gauge propagators appear in the loop.

Diagram $(g)$ contributes
\begin{equation}
\pi_{A,L} \; {\rm from} \; (g) = \frac{2}{\beta_L} \delta_{AB}
	\frac{1}{2} \int_{\left[-\pi,\pi\right]^3} 
	\frac{(\widetilde{2k})_i (\widetilde{2k})_j}
	{(\widetilde{k-p/2})^2 (\widetilde{k+p/2})^2} 
\label{fromg}
\end{equation}
in the lattice regulation and 
\begin{equation}
\pi_{A,C} \; {\rm from} \; (g) = \frac{2}{\beta_L} \delta_{AB}
	\frac{1}{2} \int_{\Re^3} \frac{2k_i 2k_j}
	{(k-p/2)^2 (k+p/2)^2} - \; {\rm UV \; divergence} 
\end{equation}
in the continuum.  The ``$-$UV divergence'' subtracts the
value at $p=0$.  Hence the $O(p^0)$ contribution is just the lattice
value, which is (using $(\widetilde{2k})_i = 2 \tilde{k}_i \cos(k_i/2)$)
\begin{equation}
\frac{\delta_{AB}}{\beta_L} \int_{\left[-\pi,\pi\right]^3} 
	\frac{4 \tilde{k}_i \tilde{k}_j 
	\cos(k_i/2) \cos(k_j/2)}{(\tilde{k}^2)^2} = 
	\frac{\delta_{AB}\delta_{ij}}{\beta_L} \int_{\left[-\pi,\pi\right]^3} 
	\frac{ 4 \tilde{k}_1^2 - \tilde{k}_1^4}{(\tilde{k}^2)^2} \, .
\label{secondp0}
\end{equation}
The integral involving $\tilde{k}_1^2$ can be performed using cubic
invariance and gives $(4/3) (\Sigma/4 \pi)$, but it is not immediately
obvious how to perform the other integral, or that its value will
correctly cancel the other $O(p^0)$ parts.  The integral is
solved using the identity, Eq. (\ref{4by4}), which was derived from
the invariance of $\int_{\left[-\pi,\pi\right]^3} \ln(\tilde{k}^2)$ 
on shifting $k_1$
(which can be compensated for by a shift in the integration range).  
It does in fact cancel the other $O(a^0)$ parts arising from scalar
loops.

To understand why, note that all 1 loop contributions to the gauge action 
from scalar loops can be computed by performing the integral over the 
Higgs fields in the Gaussian approximation (ie neglecting $\lambda_L$);
the scalar loop
contribution is $(-1/2) {\rm Tr \, ln}(-D^2+m^2)$.  This can then be
expanded about ${\rm Tr} \ln -\partial^2+m^2= \int_{\left[-\pi,\pi\right]^3} 
\ln( \tilde{k}^2 + m^2 ) $.  The required
identity came from shifting the integration variable in 
exactly this integral (at $m^2=0$), which is of course the
same as applying a spatially nonvarying gauge field.  In fact, 
Eq. (\ref{before4by4}) contains precisely the integrals in Eqs. 
(\ref{firstp0}) and (\ref{secondp0}) which contribute at $O(p^0)$.

Expanding the contribution from diagram $(g)$ 
to $O(p^2)$ one finds, from Eq. (\ref{fromg}),
\begin{eqnarray}
\frac{\delta_{AB}}{\beta_L}
\int_{\left[-\pi,\pi\right]^3} \frac{4 \tilde{k}_i \tilde{k}_j 
	\cos(k_i/2) \cos(k_j/2)}
	{(\tilde{k}^2)^2} \left[ \sum_l \frac{p_l^2}{2 \tilde{k}^2}
	\left(\frac{1}{2} \tilde{k}_l^2 - 1 \right) + \right.
\nonumber \\
	\left. \sum_{lm} p_l p_m \frac{\tilde{k}_l \tilde{k}_m
	\cos(k_l/2) \cos(k_m/2)}{(\tilde{k}^2)^2} \right] \, .
\end{eqnarray}
The corresponding continuum contribution is
\begin{equation}
\frac{\delta_{AB}}{\beta_L} \int_{\Re^3} \frac{4 k_i k_j}{k^4}
	\left[ - \frac{p^2}{2 k^2} + \frac{(p \cdot k)^2}{k^4}
	\right] \, ,
\end{equation}
which removes the infrared divergence of the lattice version.  The 
difference between the lattice and continuum expressions must be
of form
\begin{equation}
\frac{\delta_{AB}}{\beta_L}
	(A \delta_{ij} p^2 + B p_i p_j + C \delta_{ij} p_i^2) \, ,
\end{equation}
simply from cubic invariance.  We expect that the answer will be
rotationally invariant and transverse, $B = -A$ and $C=0$, because
all gauge invariant, cubic invariant, dimension 4 operators 
are; and this
will constitute a check on the calculation.  The $p_1^2$ contribution
when $i=j=1$ is
\begin{equation}
A+B+C = 4 \int_{\left[-\pi,\pi\right]^3} 
	\frac{\tilde{k}_1^2(1-\tilde{k}_1^2/4)}
	{(\tilde{k}^2)^2} \left( \frac{ \tilde{k}_1^2 - 2}{4 \tilde{k}^2}
	+ \frac{\tilde{k}_1^2(1-\tilde{k}_1^2/4)}
	{(\tilde{k}^2)^2} \right) - \; {\rm IR \; divergence} \, ,
\end{equation}
which can be evaluated using Eq. (\ref{firstby8}) and Eq. (\ref{firstby6})
in Appendix A; it vanishes.  One can find $A$ from the $p_2^2$
contribution when $i=j=1$; it equals
\begin{equation}
A = 4 \int_{\left[-\pi,\pi\right]^3} \frac{\tilde{k}_1^2(1-\tilde{k}_1^2/4)}
	{(\tilde{k}^2)^2} \left( \frac{ \tilde{k}_2^2 - 2}{4 \tilde{k}^2}
	+ \frac{\tilde{k}_2^2(1-\tilde{k}_2^2/4)}
	{(\tilde{k}^2)^2} \right) - \; {\rm IR \; divergence} \, ,
\end{equation}
which can be evaluated using Eq. (\ref{secondby8}) and Eq. (\ref{secondby6}),
and equals
\begin{equation}
A = \frac{1}{12} \frac{\Sigma}{4\pi} - \frac{1}{3} \frac{\xi}{4\pi} \, .
\end{equation}
To evaluate B one takes the $p_1p_2$ part when $i=1$ and $j=2$, which
is
\begin{equation}
B = 8 \int_{\left[-\pi,\pi\right]^3} \frac{ \tilde{k}_1^2 \tilde{k}_2^2 
	(1 - \tilde{k}_1^2/4) ( 1 - \tilde{k}_2^2/4 )}
	{ ( \tilde{k}^2 )^4} - \; {\rm IR \; divergence} = -A \, ,
\end{equation}
evaluating it with the same equations.  Hence $C=0$ and the result is
indeed rotationally invariant and transverse.

There are no new techniques involved in the remaining integrals, 
so we will simply tabulate all of the results here.

Five diagrams, $(a)$ through $(e)$ of Figure \ref{Higgs_self}, contribute
to the Higgs self-energy.  They are evaluated above, and equal
\begin{eqnarray}
\beta_L (\pi_{\Phi,L} - \pi_{\Phi,C}) & = & -6 \lambda \frac{\Sigma}{4 \pi}
	\quad {\rm from}\; (a) \quad + \\
& & ( - 6 - 3 \alpha ) \frac{\Sigma}{4 \pi} + p^2 (1 + \frac{\alpha}{2})
	\frac{\Sigma}{4 \pi} \quad {\rm from}\; (b) \quad + \\
& & 3 \alpha \frac{\Sigma}{4 \pi} + p^2 \left( (-\frac{1}{2} - 
	\frac{\alpha}{2}) \frac{\Sigma}{4 \pi} + (9 - 3 \alpha) 
	\frac{\xi}{4 \pi} \right) \quad {\rm from} \; (c) \quad + \\ 
& & \left[ ( - 2 - \alpha ) \frac{\Sigma}{4 \pi} + p^2 (
	\frac{1}{3} + \frac{\alpha}{6})
	\frac{\Sigma}{4 \pi} \right] \tan^2 \Theta_W
	\quad {\rm from}\; (d) \quad + \\
& & \left[ \alpha \frac{\Sigma}{4 \pi} + p^2 \left( (-\frac{1}{6} - 
	\frac{\alpha}{6}) \frac{\Sigma}{4 \pi} + (3 - \alpha) 
	\frac{\xi}{4 \pi} \right) \right] \tan^2 \Theta_W
	\quad {\rm from} \; (e) \, .
\end{eqnarray}

Seven diagrams, $(f)$ through $(l)$ of Figure \ref{SU2_self}, contribute to 
the SU(2) field self-energy.  In Feynman ($\alpha=1$) gauge, they 
contribute ($\delta_{AB}$ understood)
\begin{eqnarray}
\beta_L (\pi_{A,L} - \pi_{A,C}) & = & \delta_{ij}
	\left( -2 \frac{\Sigma}{4\pi} + \frac{1}{3}  \right)
	\quad {\rm from} \; (f) \quad + \\
& & \delta_{ij} \left( 2 \frac{\Sigma}{4\pi} - \frac{1}{3}  \right)
	+ (\delta_{ij} p^2 - p_i p_j ) 
	\left( \frac{1}{12} \frac{\Sigma}{4\pi} - 
	\frac{1}{3} \frac{\xi}{4\pi} \right)
	{\rm from} \; (g) \quad + \\
& & \frac{-2}{3} \delta_{ij} \qquad {\rm from} \; (h) \qquad + \\
& & \frac{-4}{9} \delta_{ij} \qquad {\rm from} \; (i) \qquad + \\
& & \delta_{ij} \left( -4 \frac{\Sigma}{4\pi} + \frac{2}{3} \right) 
	+ \delta_{ij} p^2 \left( -\frac{1}{6} \frac{\Sigma}{4\pi}
	+ \frac{2}{3} \frac{\xi}{4\pi} \right)  \nonumber \\
	+ p_i p_j  \left(
	\frac{1}{6} \frac{\Sigma}{4\pi} + \frac{4}{3} 
	\frac{\xi}{4 \pi} \right)
	\quad {\rm from} \; (j) + \\
& & \delta_{ij} \left( -16 \frac{\Sigma}{4\pi} + \frac{28}{9} \right) +
	\delta_{ij} p^2 \left( \frac{14}{3} \frac{\Sigma}{4\pi}
	\right) + \nonumber \\  & & \quad
	p_i p_j \left( -\frac{8}{3} \frac{\Sigma}{4\pi}
	- \frac{1}{3} \right) + \delta_{ij} p_i^2 
	\left( -2 \frac{\Sigma}{4\pi} + \frac{1}{3} \right)
	\quad {\rm from} \; (k) \quad +  \\
& & \delta_{ij} \left( 20 \frac{\Sigma}{4\pi} - \frac{4}{3} \right) +
	\delta_{ij} p^2 \left( 14 \frac{\xi}{4\pi} 
	- \frac{29}{6} \frac{\Sigma}{4\pi}
	+ \frac{1}{3} \right) + \nonumber \\
	& & \quad p_i p_j \left( -16 \frac{\xi}{4\pi} 
	+ \frac{17}{6} \frac{\Sigma}{4\pi}
	\right) + \delta_{ij} p_i^2 \left( 2 \frac{\Sigma}{4\pi}
	- \frac{1}{3} \right) \quad {\rm from} \; (l) \, .
\end{eqnarray}
Only the last two are gauge parameter dependent.  The total is
rotationally invariant, transverse, and free of $O(p^0)$ contributions.

\begin{figure}
\centerline{\psfig{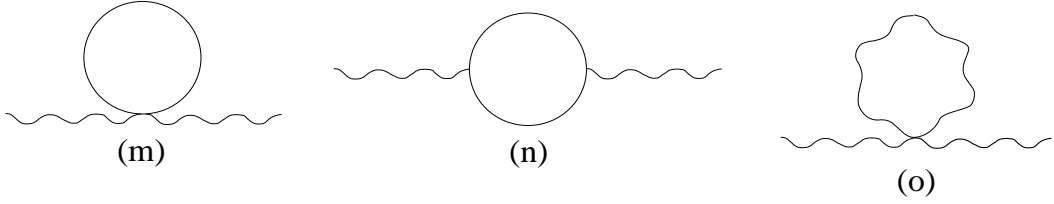}}
\caption{\label{U1_self} Contributions to the U(1) particle self-energy.
The wiggly lines are the U(1) gauge particle, and the straight lines
are the scalars.  Diagram $(o)$ only exists in the lattice theory.}
\end{figure}

Only three diagrams, $(m)$ through $(o)$ in Figure \ref{U1_self},
contribute to $\pi_B$.  Scalar loops contribute the same value as for the
SU(2) self-energy, because although the SU(2) contribution is 
enhanced by a group factor of 2, the SU(2) doublet contains 2 fields
which are charged under U(1).  The result is
\begin{eqnarray}
\frac{\beta_L}{\tan^2 \Theta_W} (\pi_{B,L} - \pi_{B,C}) & = & \delta_{ij}
	\left( -2 \frac{\Sigma}{4\pi} + \frac{1}{3}  \right)
	\quad {\rm from} \; (m) \quad + \\
& & \delta_{ij} \left( 2 \frac{\Sigma}{4\pi} - \frac{1}{3}  \right)
	+ (\delta_{ij} p^2 - p_i p_j ) \times \nonumber \\ & &  
	\left( \frac{1}{12} \frac{\Sigma}{4\pi} - 
	\frac{1}{3} \frac{\xi}{4\pi} \right)
	\quad {\rm from} \; (n) \quad + \\
& & \frac{1}{3} (\delta_{ij} p^2 - p_i p_j) \qquad {\rm from}\; (o)
\end{eqnarray}
The factor of $\tan^2 \Theta_W$ arises as follows.  
Each $B$ field propagator contributes
a $\tan^2 \Theta_W$, and there is 1 more propagator 
when a self-energy insertion occurs than when none occurs; in the
case of diagram $(o)$, there are two extra propagators, but the
vertex contributes $\tan^{-2} \Theta_W$.  Although this diagram
does not exist in the continuum theory, it completely dominates the
contribution to the self-energy.  It arises from the compact nature
of the gauge action, and its value is given {\it exactly} by the
tadpole improvement technique at one loop.

\begin{figure}
\centerline{\psfig{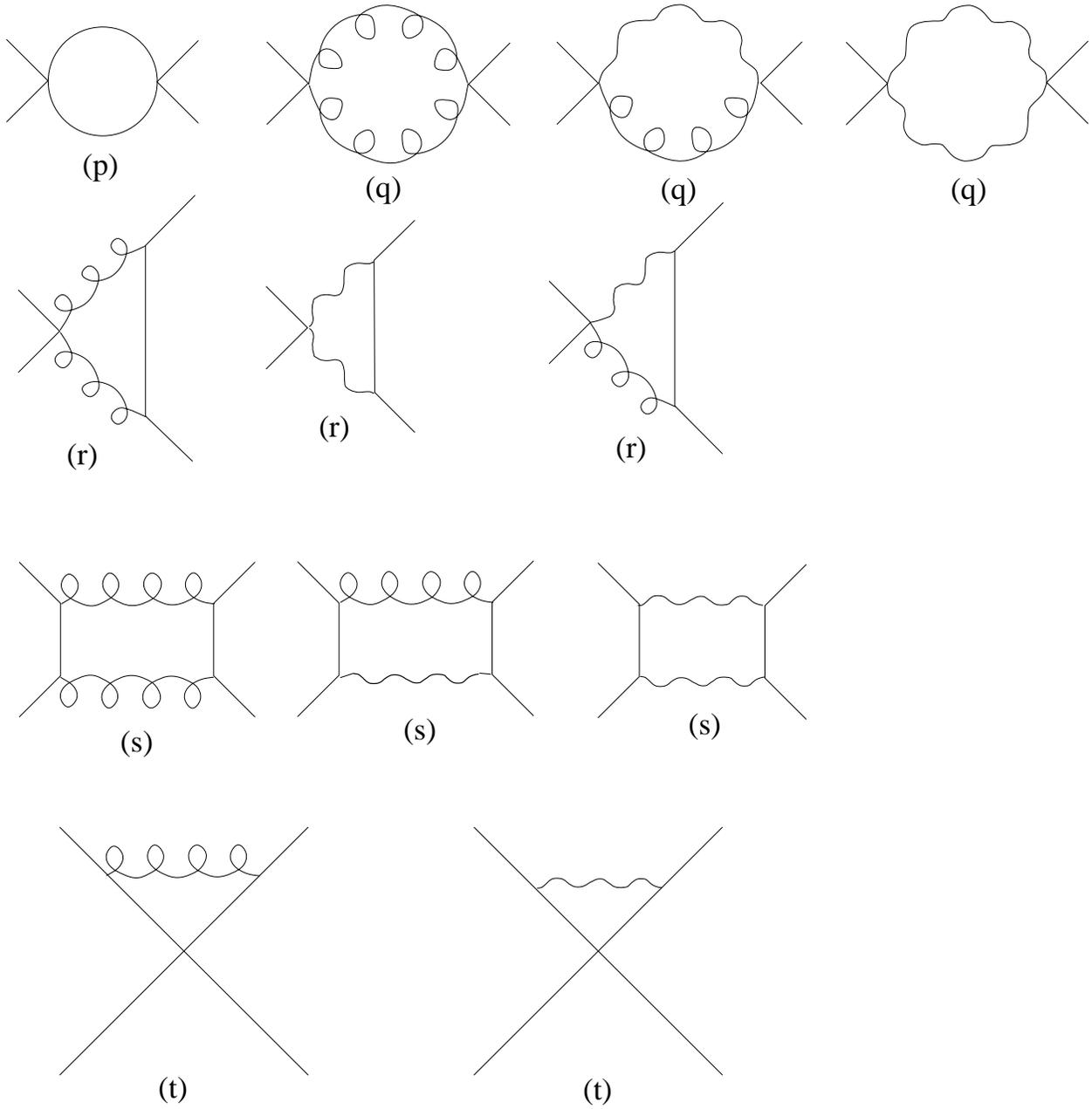}}
\caption{\label{Scalar_4} Contributions to the scalar 4-point function,
which correct $\lambda_L$.  All but $(p)$ and $(q)$ vanish in Landau
gauge.  The curly lines are SU(2) gauge propagators, the wiggly lines
are U(1) gauge propagators.}
\end{figure}

Some 12 diagrams contribute to the scalar four point vertex, in a general
gauge.  All but 4 of these vanish in Landau gauge; also the 
diagrams involving gauge lines can be grouped by topology, with
the type of gauge particle ($A$ or $B$) only determining counting
factors.  The diagrams are listed in Figure \ref{Scalar_4};
their contributions are
\begin{eqnarray}
\beta_L (V_{\lambda,L} - V_{\lambda,C}) & = & -12 \lambda^2 
	\frac{\xi}{4\pi} \quad {\rm from} \; (p) \quad + \\
& & - (4 + 2 (1 + \tan^2 \Theta_W)^2) \frac{\xi}{4\pi} + \nonumber \\
	& & - (2 + (1 + \tan^2
	\Theta_W)^2 ) \alpha^2 \frac{\xi}{4\pi} 
	\quad {\rm from} \; (q) \quad + \\
& & 2 (2 + (1 + \tan^2 \Theta_W)^2 ) \alpha^2 \frac{\xi}{4\pi}
	\quad {\rm from} \; (r) \quad + \\
& & - (2 + (1 + \tan^2 \Theta_W)^2 ) \alpha^2 \frac{\xi}{4\pi}
	\quad {\rm from} \; (s) \quad + \\
& & (6 + 2 \tan^2 \Theta_W) \alpha \frac{\xi}{4\pi} \lambda
	\quad {\rm from} \; (t) \, .
\end{eqnarray}
The $\alpha^2$ terms all cancel, and the $\alpha$ term absorbs the
$\alpha$ term in $Z_{\Phi}$ in the computation of $\delta \lambda_L$.
The Landau gauge value could have been derived by inserting $m^2$
in the propagators of the ``tadpole'' contributions to the scalar
self-energy and finding the $O(m^2)$ term, or equivalently by expanding
the (tadpole) one loop contribution to the Landau gauge effective
potential to a sufficient power in $m$, as discussed in Appendix A.

\begin{figure}
\centerline{\psfig{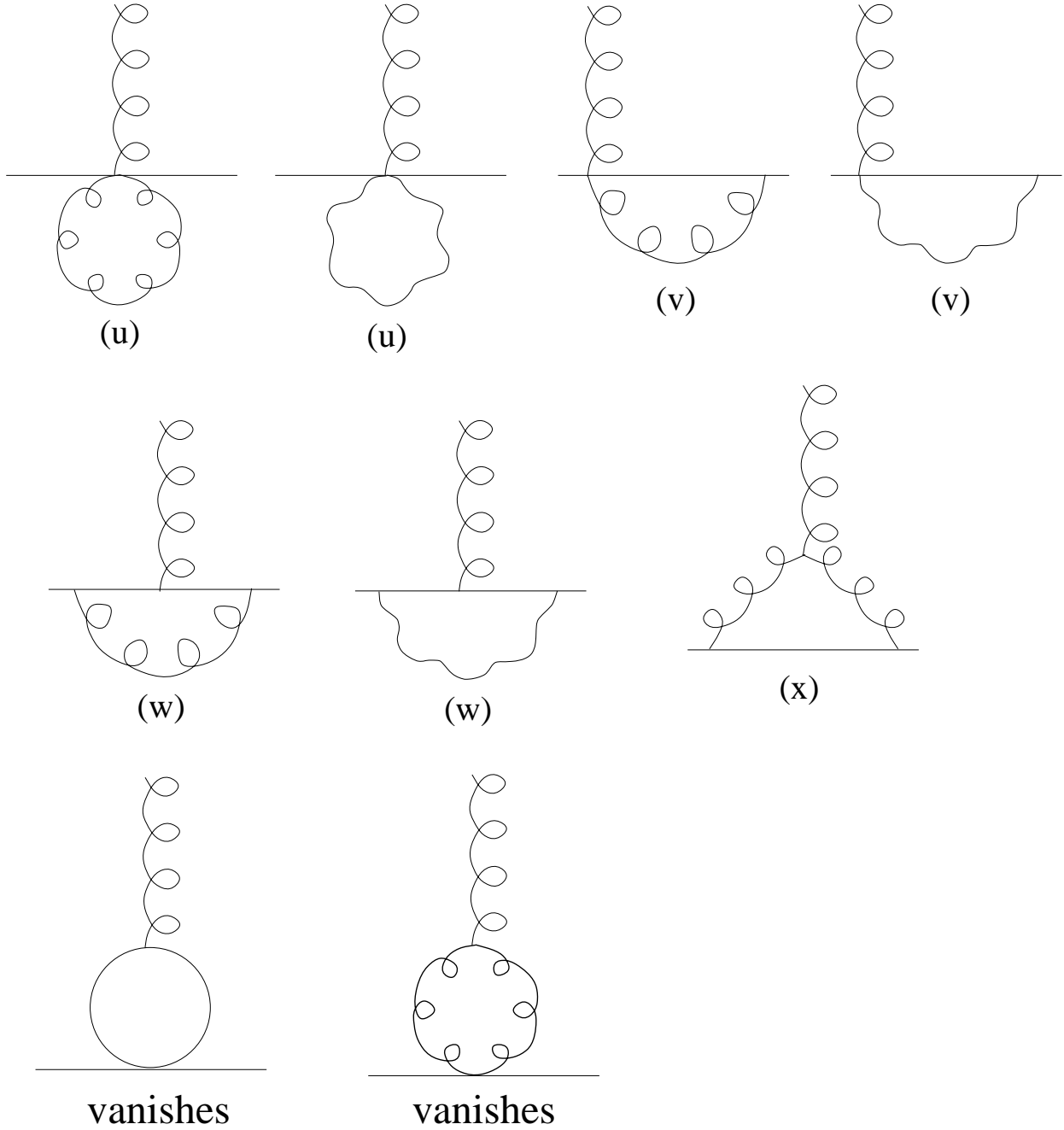}}
\caption{\label{A_vertex} Contributions to the SU(2) Higgs vertex.
The wiggly lines are U(1) propagators, the curly ones are SU(2)
propagators, the straight ones are scalar propagators.  The last two
diagrams vanish on integration over the loop momenta.}
\end{figure}

Seven diagrams contribute to the scalar-$A$ field 3 point vertex.  
Again, they can be grouped by topology; they are presented in
Figure \ref{A_vertex}.  Their contributions are
\begin{eqnarray}
\beta_L (V_{A,L} - V_{A,C}) & = & \left( \frac{5}{3} + \tan^2 \Theta_W
	\right) \left( - \frac{1}{3} - \frac{\alpha}{6} \right)
	 \frac{\Sigma}{4\pi} \quad {\rm from}\; (u) \quad + \\
& & \left( 1 + \tan^2 \Theta_W \right) \left( -3 \frac{\xi}{4\pi} 
	+ \frac{1}{6} \frac{\Sigma}{4\pi} 
	+ \frac{\alpha}{3} \frac{\Sigma}{4\pi} 
	\right) \quad {\rm from}\; (v) \quad + \\
& & \left( -1 + \tan^2 \Theta_W \right) \left( \alpha \frac{\xi}{4\pi}
	- \frac{\alpha}{6} \frac{\Sigma}{4\pi} \right) 
	\quad {\rm from} \; (w) \quad + \\
& & \left( 3 \alpha \frac{\xi}{4\pi} - \frac{\alpha}{6} \frac{\Sigma}{4\pi}
	\right) \quad {\rm from} \; (x) \, .
\end{eqnarray}
When using this result to get $Z_A$ one must remember that the 
$A$ field self-energy was only evaluated at $\alpha=1$; note that
there is a nonzero $\alpha$ dependence in $V_{A} + \pi_{\Phi}$,
and hence also in $\pi_A$.
	 
The diagrams contributing to $V_B$ are the same as $(u)$ through
$(w)$, but with $A$ and $B$ lines switched.  The contributions are
\begin{eqnarray}
\beta_L (V_{B,L} - V_{B,C}) & = & ( 3 + \tan^2 \Theta_W ) 
	\left( - \frac{1}{3} - \frac{\alpha}{6} \right)
	 \frac{\Sigma}{4\pi} \quad {\rm from}\; (u) \quad + \\
& & ( 3 + \tan^2 \Theta_W ) \left( -3 \frac{\xi}{4\pi} 
	+ \frac{1}{6} \frac{\Sigma}{4\pi} 
	+ \frac{\alpha}{3} \frac{\Sigma}{4\pi} 
	\right) \quad {\rm from}\; (v) \quad + \\
& & ( 3 + \tan^2 \Theta_W ) \left( \alpha \frac{\xi}{4\pi}
	- \frac{\alpha}{6} \frac{\Sigma}{4\pi} \right) 
	\quad {\rm from} \; (w) \, .
\end{eqnarray}
The total is precisely minus the total $O(p^2)$ contribution to the Higgs
self-energy; hence $V_B$ and $Z_{\Phi}$ cancel in the evaluation of
$Z_B$ and only $\pi_B$ contributes.  The same thing happens in 
the renormalization of continuum, 4 dimensional U(1) gauge theory.

\begin{figure}
\centerline{\psfig{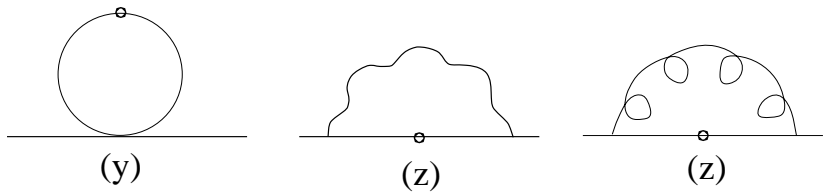}}
\caption{\label{Op_renorm} Contributions to the $\Phi^2$ operator
renormalization.  The dots are $\Phi^2$ operators, the plain lines
are scalar propagators, the curly line is the SU(2) propagator, and
the wiggly line is the U(1) propagator.  The diagrams with gauge
particles automatically vanish (at zero momentum) in Landau gauge.}
\end{figure}

Finally, there are three contributions to the multiplicative renormalization
of the $\Phi^2$ operator insertion, listed in Figure \ref{Op_renorm}.
The contributions to $L_{\Phi}$ are
\begin{eqnarray}
\beta_L (L_{\Phi,L} - L_{\Phi,C}) & = & - 6 \lambda \frac{\xi}{4\pi}
	\quad {\rm from} \; (y) \quad + \\
& & \alpha \left( 3 + \tan^2 \Theta_W \right) \frac{\xi}{4 \pi}
	\quad {\rm from} \; (z) \, ,
\end{eqnarray}
which will cancel the $\alpha$ dependence in $Z_{\Phi}$ when one forms
the gauge invariant combination $Z_{\Phi} Z_{OP}^{-1}$.

This concludes the evaluation of the relevant diagrams.

\section{Conclusion}
\label{conclusion}

We have argued that the substantial linear in $a$ corrections arising
in lattice Monte-Carlo investigations of the strength of the 
electroweak phase transition arise from the difference between 
ultraviolet screening of couplings and wave functions in the 
continuum and lattice theories, and can be cured either with
an $O(a)$ shift in the relation between the continuum and lattice
couplings and wave functions or equivalently with an $O(a)$ shift in the
interpretation of lattice data (the values of $\lambda$ and
$\beta_L$ used and the jumps in order parameters).  

\begin{figure}
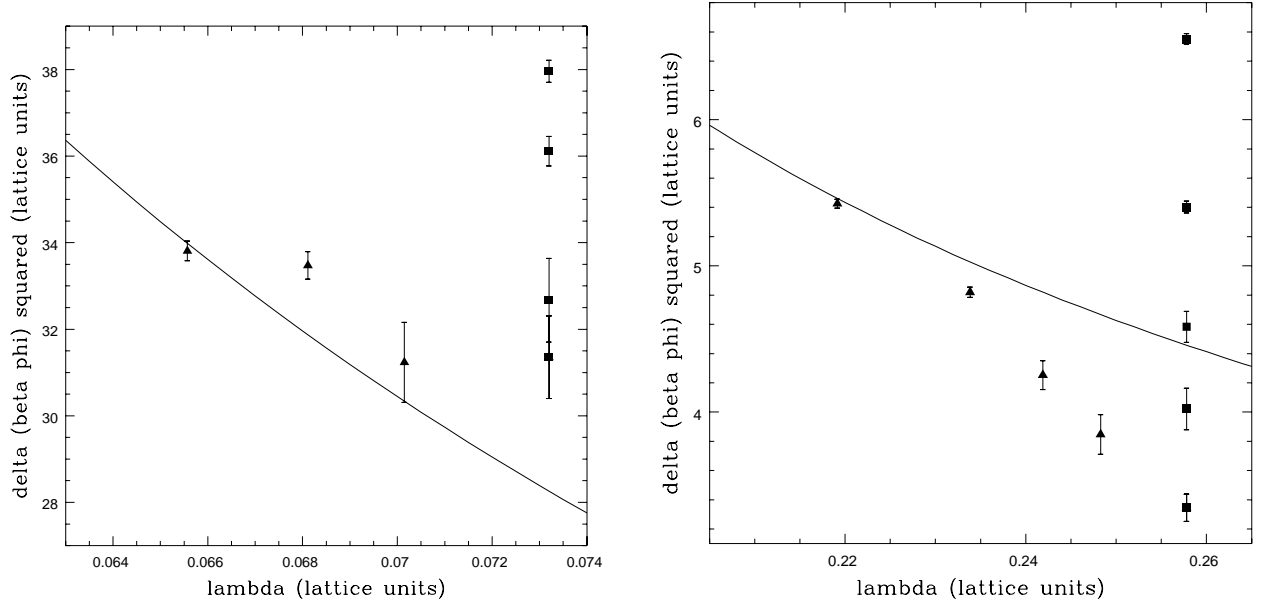

\mbox{\psfig{file=mh35.epsi,width=3.1in}}
\hspace{0.2in}
\mbox{\psfig{file=mh60.epsi,width=3.1in}}
\caption{\label{correcting} Jump in $\beta_L^2 \Phi^2$ as a function of
scalar self-coupling, before $O(a)$ corrections (squares) and after
$O(a)$ corrections (triangles).  The data is that used in 
\protect{\cite{several11}}.  The figure at the left is the $m_H^*=35$GeV
data, at $\beta_L = 8,12,20$ (from top to bottom before improvement and
from small $\lambda$ to large $\lambda$ after), and the figure at 
the right is the $m_H^* = 60$GeV data, at $\beta_L = 5,8,12,20$.  In each
case the last unimproved datapoint is the result of a linear extrapolation
of the others.}
\end{figure}

To test the technique, we re-examine existing data, the infinite volume
extrapolations of the jump in the order parameter 
$\Delta (\beta_L^2 \Phi^2)$ (the difference of $\Phi^2$ between
phases at the equilibrium temperature, extrapolated to the infinite
volume limit)
for $m^*_H = 35$ and $60$GeV, computed in
\cite{several11}.  We applied $O(a)$ corrections to the data
using Eqs. (\ref{impbeta})--(\ref{impphi})\footnote{There is
an $O(a^2)$ ambiguity involved in solving these equations--should one
compute $Z_A$ etc. using the uncorrected or corrected couplings?  We
used the corrected couplings, which means that the expressions had
to be solved iteratively.}.  We plot the uncorrected and corrected 
jumps in the order parameter in Figure \ref{correcting}, which also
shows the two loop perturbative value as a function of $\lambda$.
The corrected data tell a consistent story.  At $m^*_H=35$GeV the two loop 
perturbation theory is working fairly well, underestimating somewhat
the jump in the order parameter, presumably because of transition 
strengthening three loop effects \cite{several11} 
(though the coarsest lattice seems inconsistent with the other two, 
but see below), and at $60$GeV the jump is 
beginning to fall below the perturbative prediction.
Before the improvement the points demand a substantial extrapolation; the
result of this extrapolation is also shown in the figure.  Such an
extrapolation should successfully remove $O(a)$ errors, but
it is less optimal than improving the lattice data analytically because
the extrapolation is numerically expensive due to the demands of the finest
lattices, and the errors are dominated by the $O(a^2)$ errors in the
coarsest lattice (which may also be amplified by the extrapolation
process) and the statistical errors in the finest lattice.  Using
a single intermediate coarseness lattice and the $O(a)$ improvements 
could give smaller statistical and systematic errors at less numerical
cost.

For the $m^*_H=35$GeV data the linear extrapolation of the uncorrected
data is quite poor, with $\chi^2 = 2.9$ for 1 degree of freedom.  Comparing
the corrected data to the perturbative estimate, we see that the two
finer lattices are following the expected trend of decreasing $\phi$ with
increasing $\lambda$; but the data on the coarsest lattice buck the
trend, suggesting that the blame for the poor fit 
falls on the data from the coarsest lattice ($\beta_L = 8$).
This is probably due to $O(a^2)$ effects.  At
such a small value of $\lambda$, the phase transition is very strong,
and the particle masses in the broken phase are quite large.  The
natural scale of the physics involved in the phase transition is quite
short, short enough that one should worry about corrections from high
dimension operators.  We can estimate the size of such corrections by
seeing how much one such effect, a $\phi^5/\beta_L^2$ term
induced in the effective potential at one loop by nonrenormalizable
derivative corrections (computed in Appendix A), changes the 
perturbative value.  Its contribution to
the effective potential, summing only over the transverse gauge fields, is
\begin{equation}
\frac{6}{160 \pi \beta_L^2} \left( \Phi^2 \right)^{5/2} \, , \;
{\rm lattice \; units \, ,} \qquad {\rm or} \; \;
\frac{6}{160 \pi} \frac{g^5 a^2 T}{32} \phi^5 \, , \; {\rm physical
\; units} \, .
\end{equation}
Adding this to the 2 loop effective potential shifts the minimum (in the
lattice units used in the figure) from $\beta_L^2 \Phi^2 = 35.4$ 
to $\beta_L^2 \Phi^2 = 30.2$,
which is of the right sign and the right general size to account for
the discrepancy between the data and the perturbative value.  
(This is not the only $O(a^2)$ effect, so we should not have expected
good agreement after correction--it can only be used to estimate what
accuracy we can expect.)  The correction for the next datapoint is
already smaller by a factor of $0.4$.  Because of this error, the 
extrapolation of the unimproved data suggest a jump in  the order parameter
which is larger than the perturbative value, by an amount which is
inconsistent with the corrected data from the two finer lattices.
This is an example of the danger of linear extrapolations; $O(a^2)$ systematic
errors from the coarsest lattice data appear in the final answer,
which depends very strongly on the coarse lattice data because 
its statistical errorbars are small.

The $O(a^2)$ correction from nonrenormalizable operators is smaller for 
the $60$GeV data, because a $\phi^5$ term changes the strength of the
phase transition far more for small $\lambda$ than for large $\lambda$;
from the one loop perturbative effective potential we would estimate
the fractional change in the order parameter $\phi$ due to such
a term to be $\propto \phi/\lambda \sim 1/\lambda^2$.  A small value of
$\lambda$ demands a very fine lattice, which makes intuitive sense because
a stronger phase transition involves physics on shorter length scales.
Hence, the extrapolation of the $60$GeV data much better represents the 
trend in the improved data.

The technique we have presented here is not restricted to 
correcting the jump in the order parameter $\Phi^2$.
One can also correct computed surface tensions, by replacing
the naive values of $\beta_L$ and $\lambda$ with the corrected 
versions in those calculations.  Improving the surface tension calculation
is particularly easy because it only depends on a physical length scale
and ratios of probabilities for different values of the order 
parameter, and not on any operator insertion which might renormalize.
To remove $O(a)$ errors from
the calculation of other order parameters, such as the string bit or
the Wilson loop, one must still compute the one loop
corrections to the appropriate operator insertions.

While we estimate the size of an $O(a^2)$ error above,
so far we have said nothing about removing $O(a^2)$ errors.  These arise
not only from the finite renormalization of the bare couplings and
wave functions, but from tree level nonrenormalizable operators, which
modify the infrared behavior of the theory at $O(a^2)$ (for instance,
the $\phi^5$ term mentioned above).
To remove them one should start out with an ``improved'' action 
\cite{improvedaction}, rather than the minimal Wilson action we discuss here.
Unfortunately the ultraviolet effects computed here will be different
in such an improved action and must be recomputed.

This still leaves out the $O(a^2)$ corrections to couplings and wave
functions which arise from two loop graphs.  Directly computing these
would involve 205 two loop diagrams for SU(2) Higgs theory,
\footnote{in Landau gauge, where most of the scalar four 
point vertex corrections vanish
automatically, there are 17 diagrams contributing to this
vertex, 31 diagrams contributing to the scalar self-energy, 35 diagrams
contributing to the $\Phi^2$ operator correction, 50 diagrams 
contributing to the gauge self-energy, and 72 nonvanishing 
diagrams contributing to the gauge-scalar vertex.  
In a general gauge there are even more.} and even
more if the U(1) factor is included.  It is clearly impractical to attempt
this renormalization.  However, one can absorb the great majority of
the $O(a^2)$ corrections to wave functions and couplings in a very 
economical fashion, by using the tadpole improvement scheme.  First,
the action is tadpole improved by the technique developed by
Lepage and Mackenzie \cite{Lepage}, thereby correcting most of
the loop contributions to the wave functions and couplings, at every
order.  Then $O(a)$ corrections to the wave functions and couplings
are applied, to compensate for the difference between the full $O(a)$
corrections and those $O(a)$ corrections which the tadpole scheme will
have already made.  This removes all $O(a)$ errors and most 
$O(a^2)$ and higher order errors.

Note finally that while removing all $O(a^2)$ errors may
be impossible in practice, removing all $O(a^3)$ errors is probably
impossible even in principle, because there are $O(a^2)$ corrections
to the ultraviolet propagators, brought about by their couplings to
infrared nonperturbative physics, which are not computable and which
propagate via the one loop UV corrections discussed here into
$O(a^3)$ errors in the infrared wave functions and propagators.  However,
from a pragmatic point of view, since the dimensional reduction 
program itself has a limited accuracy, and since it is quite easy in
practice to drive $O(a^3)$ corrections below the $1\%$ level, this is
not of practical importance.

\centerline{Acknowledgements}

I would like to thank Kari Rummukainen for providing the data presented
in the conclusion, and Peter Lepage and Keijo Kajantie for useful
conversations and correspondence.

\appendix

\section{Integrals and Identities}
\label{appendixA}

In this appendix we expand the tadpole integral in powers of mass, evaluating
two of the resulting integrals numerically.  Then we derive a number of
identities which relate all other integrals needed in this paper to these
two.

The basic integral encountered in finding the effective potential to
one loop in the lattice regulation with the minimal (Wilson) action is
\cite{FKRS}
\begin{equation}
a I(m) = \int_{ \left[ -\pi , \pi \right]^3} \frac{d^3 k}{(2 \pi)^3}
\frac{1}{\tilde{k}^2 + (am)^2} \, ,
\end{equation}
where, as in the text, $\tilde{k}_i = 2 \sin(k_i/2)$ and 
$\tilde{k}^2 = \sum_{i=1,2,3} 4 \sin^2 (k_i/2)$.  Also, the 
$d^3k/(2\pi)^3$ will be assumed in all integrals in the remainder of
the section for notational simplicity.
The integral has been expanded in a power series about $am=0$ 
through $O((am)^3)$ in
\cite{FKRS}.  We repeat that expansion here, but show exactly what
integral is responsible for the third and fourth coefficients, rather
than getting them numerically by evaluating the original integral.
This is useful because the integral for one coefficient arises in
the one loop renormalization performed in the text, and the other
illustrates the influence of nonrenormalizable operators on the 
infrared physics.

The integral is well behaved about $k=0$ and so to lowest order in
$am \ll 1$ it is 
\begin{equation}
 \int_{ \left[ -\pi , \pi \right]^3} 
\frac{1}{\tilde{k}^2} = \frac{\Sigma}{4 \pi} \, , \qquad \Sigma = 3.175911536 
\, .
\label{Sigmaa}
\end{equation}
The notation is borrowed from Farakos et. al. \cite{FKRS}, who found
an analytic expression (although the value presented above is the
result of an accurate numerical integration).  This gives the (divergent)
mass squared correction.

Adding and subtracting this integral, we get
\begin{equation}
a I(m) = \frac{\Sigma}{4 \pi} - \int_{ \left[ -\pi , \pi \right]^3}
\frac{(am)^2}{ \tilde{k}^2 (\tilde{k}^2 + (am)^2)} \, .
\end{equation}
The second integral is infrared dominated, and it is best to add and
subtract the analogous continuum integral, which can be performed;
\begin{eqnarray}
\int_{ \left[ -\pi , \pi \right]^3}
\frac{(am)^2}{ \tilde{k}^2 (\tilde{k}^2 + (am)^2)} & = & 
\int_{\Re^3}  \frac{(am)^2}{k^2 (k^2 + (am)^2)}
+ \\
 & & \int_{\left[-\pi,\pi\right]^3} 
\left( \frac{(am)^2}{\tilde{k}^2 
(\tilde{k}^2 + (am)^2)} - \frac{(am)^2} {k^2(k^2 + (am)^2)} \right)
- \nonumber \\
& & \int_{\Re^3 - \left[-\pi,\pi\right]^3 } 
\frac{(am)^2}{k^2 (k^2 + (am)^2)} \, ,  \nonumber \\
\int_{\Re^3}  \frac{(am)^2}{k^2 (k^2 + (am)^2)} & = & 
\frac{am}{4 \pi} \, .
\end{eqnarray}
This gives the famous negative cubic term in the effective potential,
and leaves a remainder which vanishes as $(am)^2$.

Because $k^2 - \tilde{k}^2 = O(k^4)$ at small $k$, the remaining terms
are well behaved about $k = 0$, and we can again extract the dominant
behavior by adding and subtracting the values at $ am = 0$:
\begin{eqnarray}
& & \! \! \! \! \! \! \! 
\int_{\left[-\pi,\pi\right]^3}  
\left( \frac{(am)^2}{\tilde{k}^2 
(\tilde{k}^2 + (am)^2)} - \frac{(am)^2} {k^2(k^2 + (am)^2)} \right)
- \int_{\Re^3 - \left[-\pi,\pi\right]^3 }  
\frac{(am)^2}{k^2 (k^2 + (am)^2)} = \nonumber \\
& & \! \! \! \! \! \! \! 
 (am)^2 \left[ \int_{\left[-\pi,\pi\right]^3} 
\left( \frac{1} {(\tilde{k}^2)^2} - \frac{1}{k^4} \right) - 
\int_{ \Re^3 - \left[-\pi,\pi\right]^3}
 \frac{1}{k^4} \right] 
-  (am)^4 \times  \\
& & \! \! \! \! \! \! \! 
 \left[ \int_{\left[-\pi,\pi\right]^3} 
 \left( \frac{1}{(\tilde{k}^2)^2 (\tilde{k}^2 + (am)^2)} 
- \frac{1}{k^4( k^2 + (am)^2)} \right) - 
\int_{ \Re^3 - \left[-\pi,\pi\right]^3}
 \frac{1}{k^4(k^2 + (am)^2)} \right] 
\label{awfulpart}
\end{eqnarray}
Following the notation of \cite{FKRS} we denote the first integral as
\begin{equation} \left[ \int_{\left[-\pi,\pi\right]^3} 
	\left( \frac{1}{(\tilde{k}^2)^2} - \frac{1}{k^4} \right) - 
	\int_{ \Re^3 - \left[-\pi,\pi\right]^3}
 	\frac{1}{k^4} \right] 
	= \frac{\xi}{4 \pi} \, .
\label{Oapart}
\end{equation}
To evaluate it numerically we reduce the second integral to
a single integral over a polar angle,
\begin{equation}
\int_{\Re^3 - \left[-\pi,\pi\right]^3} 
 \frac{1}{k^4} = \frac{1}{4 \pi^3} 
+ \frac{1}{\pi^4} \int_0^{\pi/4} \cos(\theta) \arctan(\cos(\theta)) d\theta
= 0.0134035706
\end{equation}
and perform the first numerically, directly.  The result is
$\xi = 0.152859325$.  This enters the coefficient of an $O(a)$ contribution
to the effective potential which behaves as $\phi^4$; summing over
species, one can derive $V_{\lambda}$ in Landau gauge.

Finally, Eq. (\ref{awfulpart}) is infrared dominated and $O((am)^3)$,
because $k^2 - \tilde{k}^2$ is order $k^4$ at small $k$.
To extract this infrared $O((am)^3)$ piece one should expand
$\tilde{k}^2$ to $O(k^4)$,
\begin{equation}
\sum_i 4 \sin(k_i/2) = \sum_i k_i^2 - \frac{1}{12} \sum_i k_i^4 \, ,
\end{equation}
from which it follows that the contribution $-$Eq. (\ref{awfulpart}) 
to the effective potential is approximately
\begin{eqnarray}
&&(am)^4 \int_{\Re^3}  \sum_i k_i^4 \left(
\frac{2 k^2 + (am)^2}{k^4(k^2 + (am)^2)^2} - \frac{2}{k^6} \right)
\nonumber \\
& = & \frac{(am)^3}{32 \pi} \, .
\end{eqnarray}
This generates an $O(a^2)$ term in the effective potential
of form $\phi^5$ and is caused by the (tree level) nonrenormalizable
derivative terms in the action arising from the choice of lattice
regulation.  Had the expansion of $\tilde{k}^2$ been free of the $O(k^4)$
part, no such term would appear and the first one loop effect beyond
the $O(a)$ correction to $\lambda$ would be an $O(a^3)$ induced 
nonrenormalizable $\phi^6$ term.  (Of course, there would still be
corrections arising at two loops, for instance an $O(a^2)$ correction
to $\lambda$.)

The result for the tadpole integral is then
\begin{equation}
a I(m) = \frac{\Sigma}{4 \pi} - \frac{am}{4 \pi} - \frac{ \xi (am)^2}
	{4 \pi} + \frac{(am)^3}{32 \pi} + O((am)^4) \, ,
\end{equation}
and the contribution to the effective potential, $\int m I(m) dm,$ 
is
\begin{equation}
\frac{\Sigma}{4 \pi a} \frac{m^2}{2} - \frac{m^3}{12 \pi} 
	- \frac{\xi m^4 a}{16 \pi} + \frac{a^2 m^5}{160 \pi} + O(a^3m^6) \, .
\end{equation}
The first term is the linearly divergent mass squared correction.
The second is the negative cubic term which makes the phase transition
first order, the third is the contribution to the 
$O(a)$ correction $V_{\lambda}$, and the fourth is a change induced at
one loop from a nonrenormalizable derivative correction operator which
appears with $O(a^2)$ coefficient at tree level in the lattice theory.
Of these, the terms with $\Sigma$ and $\xi$ in them arise from integrals
which we need in the body of the text.

Several other integrals arise in the calculations in the text, but
a number of identities relate them all to the two (numerical) integrals, 
Eq. (\ref{Sigmaa}) and Eq. (\ref{Oapart}).  
Some of these identities involve trigonometric
manipulations of numerators, for instance using $\cos^2(k_i/2) = 1 - 
\tilde{k}_i^2/4$ and $\cos(k_i) = 1 - \tilde{k}_i^2 / 2$; we will not
bother to list these here.  Some use cubic invariance;
\begin{equation}
\int_{\left[-\pi,\pi\right]^3} 
 \frac{\tilde{k}_1^2}{(\tilde{k}^2)^2}
= \frac{1}{3} \int_{\left[-\pi,\pi\right]^3}  
\frac{\tilde{k}_1^2+ \tilde{k}_2^2 + \tilde{k}_3^2}{(\tilde{k}^2)^2}
= \frac{1}{3} \int_{\left[-\pi,\pi\right]^3} 
 \frac{1}{\tilde{k}^2}
= \frac{1}{3} \frac{\Sigma}{4 \pi} \, .
\label{2by4}
\end{equation}
Similarly,
\begin{equation}
\int_{\left[-\pi,\pi\right]^3}  
\frac{\tilde{k}_1^2}{\tilde{k}^2} = \frac{1}{3}
\label{2by2}
\end{equation}
and
\begin{equation}
\int_{\left[-\pi,\pi\right]^3}  
\frac{\tilde{k}_1^2}{(\tilde{k}^2)^3}
- \; {\rm continuum \; version} \; 
= \frac{1}{3} \frac{\xi}{4 \pi} \, .
\label{2by6}
\end{equation}
Here ``- continuum version'' means 
that $k_1^2/k^6$ should be subtracted
from the integrand and the integral of $k_1^2/k^6$ outside the first
Brillion zone should also be subtracted.
In what follows the meaning will be analogous.

Cubic invariance cannot be used to evaluate the integral over
$\tilde{k}_i^4/(\tilde{k}^2)^3$.  To do so we must take advantage of 
the periodicity of the integrand on $2\pi$ shifts in any $k_i$.  For
$\epsilon$ an infinitesimal, the periodicity of $\tilde{k}_1^2$ ensures
that
\begin{equation}
\int_{\left[-\pi,\pi\right]^3}  \frac{1}{\tilde{k}^2}  = 
\int_{\left[-\pi,\pi\right]^3}  
\frac{1}{(2 \sin^2((k_1 + \epsilon)/2))^2
+ \tilde{k}_2^2 + \tilde{k}_3^2} \, .
\end{equation} 
Now, excising a small neighborhood about the origin to avoid the
singularity there, one can expand in $\epsilon$.  At order $\epsilon^2$,
one finds after some work that
\begin{equation}
0 = \int_{\left[-\pi,\pi\right]^3 - 
{\rm excision}}  \left(
- \frac{1}{(\tilde{k}^2)^2} + \frac{1}{2} \frac{\tilde{k}_1^2}
{(\tilde{k}^2)^3} + 4 \frac{\tilde{k}_1^2}{(\tilde{k}^2)^3}
- \frac{\tilde{k}_1^4}{(\tilde{k}^2)^3} \right) + \;{\rm surface \; term}
\, , 
\label{foridentity}
\end{equation}
where the surface term arises on the boundary of the excised region.  It
cancels the linear infrared divergence occurring from the two terms
with $O(1/k^4)$ infrared behavior, so what remains is the difference
between these terms and their continuum analogs, 
eg $\int 1/(\tilde{k}^2)^2$ becomes Eq. (\ref{Oapart}).  Using
the previous integrals and identities we can integrate all but the last
term, so we find that
\begin{equation}
\int_{\left[-\pi,\pi\right]^3}  
\frac{\tilde{k}_1^4}{(\tilde{k}^2)^3}
= \frac{1}{6} \frac{\Sigma}{4 \pi} + \frac{1}{3} \frac{\xi}{4 \pi} \, .
\label{4by6}
\end{equation}
Because $ \tilde{k}_1^2 + \tilde{k}_2^2 + 
\tilde{k}_3^2 = \tilde{k}^2$, it follows from this integral, 
Eq. (\ref{2by4}), and cubic invariance that
\begin{equation}
\int_{\left[-\pi,\pi\right]^3}  
\frac{\tilde{k}_1^2 \tilde{k}_2^2}
{(\tilde{k}^2)^3}
= \frac{1}{12} \frac{\Sigma}{4 \pi} - \frac{1}{6} \frac{\xi}{4 \pi} \, .
\label{22by6}
\end{equation}

Repeating the above technique, but starting with the integral over
$\ln(\tilde{k}^2)$ rather than $1/\tilde{k}^2$, one finds that
\begin{equation}
0 = \int_{\left[-\pi,\pi\right]^3} 
\frac{-2\tilde{k}^2+\tilde{k}^2 \tilde{k}_1^2 + 4 \tilde{k}_1^2
-\tilde{k}_1^4}{(\tilde{k}^2)^2} \, , 
\label{before4by4}
\end{equation}
or
\begin{equation}
\int_{\left[-\pi,\pi\right]^3}  
\frac{\tilde{k}_1^4}{(\tilde{k}^2)^2}
= \frac{1}{3} - \frac{2}{3} \frac{\Sigma}{4 \pi} \, ,
\label{4by4}
\end{equation}
and hence also
\begin{equation}
\int_{\left[-\pi,\pi\right]^3}  
\frac{\tilde{k}_1^2 \tilde{k}_2^2}
{(\tilde{k}^2)^2}
= \frac{1}{3} \frac{\Sigma}{4 \pi} \, .
\label{22by4}
\end{equation}

Continuing the expansion to fourth order in $\epsilon$, and using the
previously derived identities, one obtains after considerable algebra that
\begin{eqnarray}
\int_{\left[-\pi,\pi\right]^3}  \frac{ \tilde{k}_1^4 
(1 - \tilde{k}_1^2/4)^2}{(\tilde{k}^2)^4} - \; {\rm continuum \; version} -
&&
\nonumber \\
\int_{\left[-\pi,\pi\right]^3}  \frac{\tilde{k}_1^2 
(1 - \tilde{k}_1^2/2) (1 - \tilde{k}_1^2/4)}{(\tilde{k}^2)^3}
- \; {\rm continuum \; version} & = & -\frac{1}{48} + \frac{1}{8}
\frac{\Sigma}{4\pi} - \frac{1}{8} \frac{\xi}{4\pi} \, .
\label{firstby8}
\end{eqnarray}
Shifting the $k_2$ integral by $\delta$, the condition that
the $O(\epsilon^2 \delta^2)$ term should vanish gives that
\begin{eqnarray}
3\int_{\left[-\pi,\pi\right]^3}  
\frac{ \tilde{k}_1^2 \tilde{k}_2^2
(1 - \tilde{k}_1^2/4)(1 - \tilde{k}_2^2/4)}
{(\tilde{k}^2)^4} - \; {\rm continuum \; version} -
&&
\nonumber \\
\int_{\left[-\pi,\pi\right]^3}  \frac{\tilde{k}_1^2 
(1 - \tilde{k}_2^2/2) (1 - \tilde{k}_1^2/4)}{(\tilde{k}^2)^3}
- \; {\rm continuum \; version} & = &  \frac{1}{32}
\frac{\Sigma}{4\pi} - \frac{1}{8} \frac{\xi}{4\pi} \, .
\label{secondby8}
\end{eqnarray}

By shifting $k_1$ by $\epsilon$ in the integral
\begin{equation}
\int_{\left[-\pi,\pi\right]^3}  
\frac{\tilde{k}_2^2}{\tilde{k}^2}
\end{equation}
and expanding to second order in $\epsilon$, one finds using 
Eq. (\ref{22by4}) that
\begin{equation}
\int_{\left[-\pi,\pi\right]^3}  \frac{\tilde{k}_1^4 
\tilde{k}_2^2 }{ (\tilde{k}^2)^3 } = \frac{1}{6} \frac{\Sigma}{4 \pi}
- \frac{2}{3} \frac{\xi}{4 \pi} \, ,
\end{equation}
and hence 
\begin{equation}
\int_{\left[-\pi,\pi\right]^3}  \frac{\tilde{k}_1^6}
{ (\tilde{k}^2)^3 } = \frac{1}{3} -  \frac{\Sigma}{4 \pi}
+ \frac{4}{3} \frac{\xi}{4 \pi} \, ,
\end{equation}
by applying the same technique used to get Eq. (\ref{22by6}) and
using Eq. (\ref{4by4}).  With these integrals we can derive that
\begin{equation}
\int_{\left[-\pi,\pi\right]^3}  \frac{\tilde{k}_1^2 
(1 - \tilde{k}_1^2/2) (1 - \tilde{k}_1^2/4)}{(\tilde{k}^2)^3}
- \; {\rm continuum \; version} = \frac{1}{24} - \frac{1}{4}
\frac{\Sigma}{4\pi} + \frac{1}{4} \frac{\xi}{4\pi} \, ,
\label{firstby6}
\end{equation}
and
\begin{equation}
\int_{\left[-\pi,\pi\right]^3}  \frac{\tilde{k}_1^2 
(1 - \tilde{k}_2^2/2) (1 - \tilde{k}_1^2/4)}{(\tilde{k}^2)^3}
- \; {\rm continuum \; version} =  - \frac{1}{16}
\frac{\Sigma}{4\pi} + \frac{1}{4} \frac{\xi}{4\pi} \, .
\label{secondby6}
\end{equation}
Hence we know the value of each integral in Eqs. (\ref{firstby8})
and (\ref{secondby8}) separately.

No other identities are needed for to complete the 1 loop renormalization.

\section{Feynman rules}
\label{appendixB}

Most of the Feynman rules needed in the calculation appear in 
\cite{Rothe}; the measure insertion (diagram $(h)$ in Section \ref{details})
is smaller by a factor of $2/3$ from the one printed there, which is
for SU(3) rather than SU(2), and similarly in the four point vertex 
in Eq. (14.44) one must set $d_{ABC}$ to 0 and the $(2/3)$ in front of 
$\delta_{AB} \delta_{BC} + \delta_{AC} \delta_{BD} + \ldots$
should become $1$.  Our conventions for the scale and the 
gauge wave function mean $a$ should be replaced by $1$ and $g$ by $2$
everywhere.  Also, the sign for the ghost coupling to two gauge
particles is backwards there.

What remains are the scalar propagator and the vertices involving the
scalar.  The propagator, in the ultraviolet where $m$ is negligible, 
is just $\delta_{ab}/p^2$.  The 4-point vertex is 
\begin{equation}
- 2 \lambda ( \delta_{ab} \delta_{cd} + \delta_{ac} \delta_{bd} 
+ \delta_{ad} \delta_{bc} )
\end{equation}
where $a,b,c,d$ are the species types of the incoming lines.  The
vertices involving gauge particles, and the pure U(1) four point
vertex, are given in Figures
\ref{feynman1} to \ref{feynman3}.  The notation is that $(i)_{ab}$ means
the action of $i$ on the species type, eg species type 1 becomes type
2 and type 2 becomes $-$ type 1, etc.  Any $i$ by itself is the imaginary
number which arises when one Fourier transforms an odd function.  The
vertices in Figure \ref{feynman3} do not have continuum analogs.  Their
role in generating extra tadpole diagrams is quite important on the
lattice; for instance they dominate the corrections $V_B$ to the U(1)-
Higgs vertex.

\begin{figure}
\centerline{\psfig{file=feynman1.epsi,width=5.7in}}
\caption{\label{feynman1} Feynman rules for the U(1) four point
function and the three point gauge-Higgs couplings.}
\end{figure}

\begin{figure}
\centerline{\psfig{file=feynman2.epsi,width=\hsize}}
\caption{\label{feynman2} Feynman rules for the four point 
gauge-Higgs couplings.}
\end{figure}

\begin{figure}
\centerline{\psfig{file=feynman3.epsi,width=\hsize}}
\caption{\label{feynman3} Feynman rules for the five 
point gauge-Higgs couplings.}
\end{figure}

\pagebreak

\section{Results in other theories}
\label{appendixC}

Here we present results for the renormalization of the abelian Higgs
model with $N$ scalars, and for SU(2)$\times$U(1) theory with adjoint
scalars (the Weinberg-Salaam model after dimensional reduction but
before integrating out heavy modes).

The action of the 3-D abelian Higgs model on a lattice is
\begin{eqnarray}
\sum_x \left[
	Z_B \sum_{i<j} ( 1 - {\rm Re} \: p_{ij} )
	+ Z_{\Phi} \sum_{b=1}^N \sum_i 
	\left( \Phi_{b}^2(x) - \Phi_{ab}(x)  u_{iac}(x)
	\Phi_{cb}(x+i) \right) + \right. \nonumber \\ \left. 
	Z_{\Phi} \frac{m_{HL}^2}{2} (\sum_b \Phi_b^2)
	+ Z_{\Phi}^2 \frac{\lambda_L}{4} (\sum_b \Phi^2_b)^2 \right] \, .
\end{eqnarray}
Here $a$ and $c$ are indicies over the real and imaginary part, which
are implicitly summed over when $\Phi^2$ appears, and $b$
is an index over the $N$ scalar species, which are assumed to have the
same mass squared and 
an $SU(N)$ symmetric quartic interaction.  The finite $a$ renormalization
only involves diagrams computed already in the text; the results are
\begin{eqnarray}
\beta_L \delta \lambda_L & = & \frac{\lambda_L}{3} \frac{\Sigma}{4\pi} + 
	 \left( 6 \lambda_L - (8 + 2 N) \lambda_L^2 - 2 \right)
	\frac{\xi}{4 \pi} \, , \\
\beta_L (Z_B - 1) & = & \frac{1}{3} + N \left( \frac{1}{24} \frac{\Sigma}
	{4 \pi} - \frac{1}{6} \frac{\xi}{4 \pi} \right) \, , \\
\beta_L \delta m^2_{HL} & = & \left(2 + (2 + 2N) \lambda_L) \right)
	\frac{\Sigma}{4 \pi} \, , \\
\beta_L ( Z_{\Phi} Z_{OP}^{-1} - 1 ) & = & \left( 3 - (2 + 2 N) \lambda_L)
	\right) \frac{\xi}{4 \pi} + \frac{1}{6} \frac{\Sigma}{4 \pi} \, .
\end{eqnarray}

Next we consider SU(2)$\times$U(1) theory, with an adjoint 
SU(2) scalar $A_0$ and an adjoint U(1) scalar $B_0$.  
If one is interested in very high precision calculations of the
electroweak phase transition strength one must include these, because
the integration over these fields (the so called heavy
modes) is lower precision than the integration over the nonzero 
Matsubara frequencies (the superheavy modes).

The integration over the nonzero Matsubara frequencies generates
mass terms for these modes.  Interaction terms already
exist at tree level, and new ones are generated at one loop in
the dimensional reduction.  The new mass terms will be 
denoted $m_A^2$ and $m_B^2$, and the new interaction terms in
the Lagrangian are
\begin{equation}
\frac{\lambda_A}{4} (A_0^2)^2 + \frac{\lambda_B}{4} (B_0^2)^2
+ \frac{h_A}{4} \Phi^2 A_0^2 + \frac{h_B}{4} \Phi^2 B_0^2
+ \frac{h_{AB}}{4} A_0^2 B_0^2 + \frac{h'}{2} B_0
\Phi_a (-A_0 \cdot \tau)_{ab} \Phi_b \, .
\end{equation}
In the minimal standard model, at lowest order in $\alpha_W$,
the bare values for the new couplings, in lattice units, are \cite{KLRS}
\begin{eqnarray}
\lambda_{AL} & = & 0 \, , \\
\lambda_{BL} & = & 0 \, , \\
h_{AL} & = & 2 \, , \\
h_{BL} & = & 2 \tan^2 \Theta_W \,  , \\
h_{ABL} & = & 0 \, , \\
h'_L & = & 2 \tan \Theta_W \, ,
\end{eqnarray}
but in what follows we will treat the general problem in which their
values are arbitrary.  Computing the $O(a)$ renormalizations involves
no topologically new diagrams, only combinatorics.  The
couplings and wave functions presented in Section \ref{setup}
are modified by
\begin{eqnarray}
\frac{1}{3} \frac{\Sigma}{4 \pi} - \frac{4}{3} 
	\frac{\xi}{4 \pi} & {\rm added \; to} &
	\beta_L (Z_A - 1) \, , \\
 - \frac{3}{4} h_{AL}^2 - \frac{1}{4}
	h_{BL}^2 - \frac{1}{2} h{'}_{L}^{2}  & {\rm added \; to} &
	\beta_L \delta \lambda_L\, , \\
 \frac{3}{2} h_{AL} + \frac{1}{2} h_{BL} 
	 & {\rm added \; to} & \beta_L \delta m_{HL}^2 \, .
\end{eqnarray}
The corrections to $Z_B$ and $Z_{\Phi}$ vanish.  

We denote the new wave function corrections as 
$Z_{A0}$ and $Z_{B0}$; the notations for the coupling corrections are
obvious.  The new couplings and wave functions renormalize by
\begin{eqnarray}
\beta_L ( Z_{A0} - 1 ) & = & 24 \frac{\xi}{4 \pi} + \frac{4}{3} 
	\frac{\Sigma}{4 \pi} - 5 \lambda_{AL} \frac{\xi}{4 \pi} \, , \\
\beta_L ( Z_{B0} - 1 ) & = & - 3 \lambda_{BL} \frac{\xi}{4 \pi} \, , \\
\beta_L \delta m_{AL}^2 & = & \left( 16 + 5 \lambda_L + 2 h_{AL} + \frac{1}{2}
	h_{ABL} \right) \frac{\Sigma}{4 \pi} \, , \\
\beta_L \delta m_{BL}^2 & = & \left( 3 \lambda_{BL} + 2 h_{BL} + \frac{3}{2}
	h_{ABL} \right) \frac{\Sigma}{4 \pi} \, , \\
\beta_L \delta \lambda_{AL} & = & - \left( 32 + 11 \lambda_{AL}^2 + h_{AL}^2 +
	\frac{1}{4} h_{ABL}^2 \right) \frac{\xi}{4\pi} + \lambda_{AL}
	\left( 48 \frac{\xi}{4 \pi} + \frac{8}{3} \frac{\Sigma}{4\pi}
	\right) \, , \\
\beta_L \delta \lambda_{BL} & = & - \left( 10 \lambda_{BL}^2 +
	h_{BL}^2 + \frac{3}{4} h_{ABL}^2 \right) \frac{\xi}{4\pi} \, , \\
\beta_L \delta h_{AL} & = & - \left( 8 + 6 \lambda_L h_{AL}
	+ 5 \lambda_{AL} h_{AL} + 2 h^2_{AL} + \frac{1}{2} h_{BL} h_{ABL}
	+ 2 h'^2_L
	\right) \frac{\xi}{4 \pi} + \nonumber \\
	& & h_{AL} \left( (33+3 \tan^2 
	\Theta_W)  \frac{\xi}{4\pi}
	+ (\frac{11}{6}+ \frac{1}{6} \tan^2 \Theta_W)
	 \frac{\Sigma}{4\pi} \right) \, , \\
\beta_L \delta h_{BL} & = & - \left( 6 \lambda_L h_{BL} + 3 
	\lambda_{BL} h_{BL} + 2 h_{BL}^2 + \frac{3}{2} h_{AL} h_{ABL}
	+ 2 h{'}_{L}^{2} \right) + \nonumber \\ & &
	 h_{BL} \left( (9+3 \tan^2 \Theta_W) 
	\frac{\xi}{4 \pi} + (\frac{3 + \tan^2 \Theta_W}{6} )
	\frac{\Sigma}{4 \pi} \right) \, , \\
\beta_L \delta h_{ABL} & = & - \left( 5 \lambda_{AL} h_{ABL} + 3 \lambda_{BL}
	h_{ABL} + 2 h^2_{ABL} + 2 h_{AL} h_{BL} + 2 h'^2_{L} 
	\right) + \nonumber \\ & & 
	h_{ABL} \left( 24 \frac{\xi}{4\pi} + \frac{4}{3}
	\frac{\Sigma}{4\pi} \right) \, , \\
\beta_L \delta h'_L & = & h'_L \left( \left( -2 \lambda_L - h_{AL}
	- h_{BL} - h_{ABL} + 21 + 3 \tan^2 \Theta_W \right)
	\frac{ \xi}{4\pi} + \right. \nonumber \\ & & \left.
	\frac{7 + \tan^2 \Theta_W}{6} 
	\frac{\Sigma}{4 \pi} \right) \, .
\end{eqnarray}
Here the wave functions already include corrections to make the
operator insertions have their naive normalization.
In fact the operator insertions will be mixed by a matrix with
$O(a)$ off diagonal terms; what we use above are the diagonal components
of the matrix.  The off diagonal terms would only be important if
the $A_0$ or $B_0$ fields took on significant condensates at the
phase transition, which they should not, because in the realistic 
case they should be given substantial (Debye) masses.  They would also
be important if we were interested in the jumps in these condensates,
but this is also not of interest.  We have not
computed the off diagonal terms here.

\end{document}